\documentclass[a4paper,11pt]{article}

\pdfoutput=1

\usepackage{graphics}%
\usepackage[mathcal]{euscript}
\usepackage[latin1]{inputenc}
\usepackage{latexsym}
\usepackage{graphicx}   
\usepackage{verbatim}  
\usepackage{color}      
\usepackage{subfigure}  
\usepackage{bm}
\usepackage{graphicx}
\usepackage[normalem]{ulem}
\usepackage{amssymb}
\usepackage{bbm}
\usepackage{float}
\usepackage{slashed}
\usepackage{amsfonts}
\usepackage{euscript}
\usepackage{amsmath}    
\usepackage{mathrsfs} 
\usepackage{pifont}
\usepackage{cancel}
\usepackage{jcappub}
\usepackage[force]{feynmp-auto}
\usepackage{comment}
\usepackage{soul}

\begin{document}%

\title{Minimally modified theories of gravity: a playground for testing the uniqueness of general relativity}


\author{ Ra{\'u}l Carballo-Rubio,}
\author{Francesco Di Filippo and}
\author{Stefano Liberati}
\affiliation{SISSA, International School for Advanced Studies, Via Bonomea 265, 34136 Trieste, Italy}
\affiliation{INFN Sezione di Trieste, Via Valerio 2, 34127 Trieste, Italy}
\emailAdd{raul.carballorubio@sissa.it}
\emailAdd{francesco.difilippo@sissa.it}
\emailAdd{liberati@sissa.it}


\abstract{In a recent paper \cite{Lin2017}, it was introduced a new class of gravitational theories with two local degrees of freedom. The existence of these theories apparently challenges the distinctive role of general relativity as the unique non-linear theory of massless spin-2 particles. Here we perform a comprehensive analysis of these theories with the aim of (i) understanding whether or not these are actually equivalent to general relativity, and (ii) finding the root of the variance in case these are not. We have found that a broad set of seemingly different theories actually pass all the possible tests of equivalence to general relativity (in vacuum) that we were able to devise, including the analysis of scattering amplitudes using on-shell techniques. These results are complemented with the observation that the only examples which are manifestly not equivalent to general relativity either do not contain gravitons in their spectrum, or are not guaranteed to include only two local degrees of freedom once radiative corrections are taken into account. Coupling to matter is also considered: we show that coupling these theories to matter in a consistent way is not as straightforward as one could expect. Minimal coupling, as well as the most straightforward non-minimal couplings, cannot be used. Therefore, before being able to address any issues in the presence of matter, it would be necessary to find a consistent (and in any case rather peculiar) coupling scheme.
}

\maketitle

\def\HRULE{{\bigskip\hrule\bigskip}}

\section{Introduction}
The authors of \cite{Lin2017} have discussed the existence of a large class of theories that have the same number of degrees of freedom as general relativity. The basic fields that are used in order to construct these theories are the 3-dimensional metric $h_{ij}$, the lapse $N$ and the shift $N^i$. While this corresponds to the familiar ADM decomposition in general relativity, in which the spacetime metric reads
\begin{equation}
\text{d}s^2=-N^2\text{d}t^2+h_{ij}(\text{d}x^i+N^i\text{d}t)(\text{d}x^j+N^j\text{d}t),
\end{equation}
these theories deviate from general relativity in the imposition of invariance under spatial diffeomorphisms only. Of course, one could naively expect that breaking full diffeomorphism invariace should end up introducing extra degrees of freedom. Hence, in order to maintain the number of degrees of freedom fixed, it is necessary to include other gauge symmetries that remove the possible additional degrees of freedom. In \cite{Lin2017} it was shown that this is indeed possible for a large class of theories, in spite of these being generally higher-derivative theories.
Let us stress that we will be dealing only with local field theories, in the sense that the corresponding Lagrangian densities are functions of the fields and their derivatives evaluated in a single point in spacetime. If one relaxes this condition, it is possible to find non-trivial extensions of general relativity \cite{Cortes2015,Cortes2016}.

A central question is whether or not these theories are equivalent to general relativity. It is possible that, in spite of being apparently different, these are however physically equivalent (but are expressed in terms of a non-standard choice of variables). Here we would like to understand this question. As we will see, clarifying this point has its subtleties due to the non-linear nature of general relativity. In Sec. \ref{sec:hamfam}, starting directly with the Hamiltonian picture, we will discuss the non-linear algebra of constraints as well as their linearization (focusing on the recovery of the Poincar\'e group in the linear limit) of a certain set of theories that includes some of the particular examples given in \cite{Lin2017}. This discussion is also presented as a warm up for the core of the paper. Then, in Sec. \ref{sec:sqrt}, we will connect the results obtained in the Hamiltonian perspective with the analysis of the Lagrangian formulation of a broad set of theories, illustrating that theories that recover the Poincar\'e group in the linear limit describe precisely gravitons (massless spin-2 particles) in this limit, as one would expect. This observation, and the fact that the 3-point amplitudes of gravitons in these examples are the same as in general relativity (as shown in detail), represent a strong hint that these theories are classically equivalent to general relativity. This conclusion is strengthened with the analysis of $n$-point amplitudes for $n\geq4$, using on-shell techniques. In Sec. \ref{sec:diff_theor} we consider, as a counterpoint, two examples of theories that are not equivalent to general relativity: in one of them the algebra of constraints does not reproduce the Poincar\'e algebra in the linear limit and, in the other one, gravitons display modified dispersion relations (and additional degrees of freedom can arise due to radiative corrections). The last part of the paper, Sec. \ref{sec:matter}, deals with additional issues that come up when couplings to matter are introduced. 

\section{A class of Hamiltonian theories \label{sec:hamfam}}

As a way to illustrate in a simpler way the main result in \cite{Lin2017}, and make the present discussion self-consistent, let us consider a class of Hamiltonian theories that are in principle structurally different from general relativity, but that still contain two local degrees of freedom. While not encompassing all the possible theories found in \cite{Lin2017}, this is however a large enough set to represent a convenient framework for our discussion (let us note that, in particular, it includes general relativity as a particular case).

The class of theories we introduce in this section are constructed from the variables $\{N,N^i,h_{ij}\}$ and their conjugated momenta $\{\pi_N,\pi_i,\pi^{ij}\}$, and are defined by the Hamiltonian
\begin{equation}\label{gen_ham}
H=\int d^3x\left(N\mathcal{H}+N^i \mathcal{D}_i+\lambda_N\pi_N+\lambda^i\pi_i\right),
\end{equation}
where
\begin{equation}
\mathcal{H}=\sqrt{h}\,F\left(R+\lambda\Pi/h\right),\qquad\qquad \mathcal{D}_{i}=-2\nabla_j\pi^j_{\ i}.\label{eq:ham&mom}
\end{equation}
Let us clarify some notation: $h$ is the determinant of the 3-dimensional Riemannian metric $h_{ij}$, $R$ is the 3-dimensional Ricci scalar associated with this very same metric (note that we are not following the usual notation $^{(3)}R$ as we will primarily work with this quantity instead of its 4-dimensional counterpart, so that there is no risk of confusion), $\Pi=\pi_{ij}\pi^{ij}-\pi^2/2$ (where $\pi=\pi^{ij}h_{ij})$, and $\nabla_i$ is the covariant derivative associated with $h_{ij}$.

Just to connect with a familiar example, let us recall that general relativity corresponds to the choice of the function $F(x)=x$ and $\lambda=1$, namely $\mathcal{H}=\sqrt{h}R+\Pi/\sqrt{h}$. For general relativity, we can easily identify the familiar expressions that appear in the so-called Hamiltonian and diffeomorphism constraints.

In all these theories, $\lambda_N$ and $\lambda^i$ are Lagrange multipliers that enforce the primary constraints
\begin{equation}
\pi_N=0,\qquad\qquad \pi_i=0,
\end{equation}
which in turn induce the secondary constraints
\begin{equation}
\mathcal{H}=0, \qquad\qquad \mathcal{D}_{i}=0.
\end{equation} 
In the particular case of general relativity, the latter correspond to the Hamiltonian and diffeomorphism constraints, respectively. We will keep using this nomenclature for any theory with total Hamiltonian given by Eq. \eqref{gen_ham}.

As in general relativity, a great deal of information is encoded in the off-shell algebra of Poisson brackets of these constraints. Following the usual practice, let us define the smeared constraints
\begin{equation}
\mathcal{H}[\alpha]=\int\text{d}^3x\,\alpha\mathcal{H},\qquad\qquad\mathcal{D}[\alpha^i]=\int\text{d}^3x\,\alpha^i\mathcal{D}_i.
\end{equation}
It follows then that (see App. \ref{sec:app1})
\begin{align}\label{gen_algebra}
&\left\{\mathcal{H}[\alpha],\mathcal{H}[\beta]\right\}=\mathcal{D}\left[\lambda F'(R+\lambda\Pi/h)^2h^{ij}\left(\beta\partial_j\alpha-\alpha\partial_j\beta\right)\right],\nonumber\\
&\left\{\mathcal{D}[\alpha^i],\mathcal{H}[\alpha]\right\}=\mathcal{H}\left[\mathcal{L}_{\alpha^i}\alpha\right],\nonumber\\
&\left\{\mathcal{D}[\alpha^i],\mathcal{D}[\beta^j]\right\}=\mathcal{D}\left[\mathcal{L}_{\alpha^i} \beta^j\right].
\end{align}
Note that $F'(x)$ denotes the derivative of $F(x)$ with respect to its single argument $x$. Several observations can be made:
\begin{itemize}
\item{The algebra is closed and all the constraints are first class, which leaves $(20-2\times 8)/2=2$ degrees of freedom. In other words, all these theories satisfy the conditions discussed in \cite{Lin2017}.}
\item{The algebra obtained is a slight modification of the usual Dirac algebra of general relativity \cite{Thiemann2008}. In particular, the only difference appears in the bracket between Hamiltonian constraints, where the structure function inside the Hamiltonian constraint is different; let us recall that for general relativity one has $\lambda F'(R+\lambda\Pi/h)^2=1$. Let us mention that similar modifications have been observed to occur in loop quantum gravity \cite{Bojowald2008,Bojowald2011,Bojowald2012,Bojowald2010,Cailleteau2011,BenAchour2016}, which makes part of our discussion below of possible interest for these studies (note that, however, signature change \cite{Bojowald2015,Bojowald2016} is not structurally allowed in the theories we are considering in this section).}
\end{itemize}
In summary, we have a large class of theories that contain two degrees of freedom, the same as general relativity, but which seem to be structurally different from the latter theory. In particular, the off-shell algebra is different from the usual Dirac algebra. One may jump to the conclusion that these theories are different from general relativity. But is this true?

Before diving into this question, let us make a remark that will be important later: the value of $\lambda$ can be changed arbitrarily by a positive multiplicative factor. This can be noticed by considering the field redefinition
\begin{equation}
\pi^{ij}\rightarrow \kappa \pi^{ij},\qquad\qquad N^i\rightarrow N^i/\kappa,
\end{equation}
with $\kappa\in\mathbb{R}$ a constant. It is straightforward to check that the combination $N^i\mathcal{D}_i$ is invariant under this rescaling, so that diffeomorphism constrain is unchanged, but that the Hamiltonian constraint becomes
\begin{equation}
\mathcal{H}=\sqrt{h}\,F\left(R+\lambda k^2\Pi/h\right)=\sqrt{h}\,F\left(R+\lambda'\Pi/h\right),
\end{equation}
with $\lambda'=\lambda \kappa^2$. Hence, it is possible to change the absolute value of $|\lambda|$ arbitrarily; note however that its sign cannot be modified by this redefinition. So, being always possible to set $\lambda=\pm 1$, it is $F(x)\neq x$ the real difference with general relativity.

\subsection{Poincar\'e algebra} \label{sec:poincare}

As a first partial answer to the question asked at the end of the previous section, we now show that in the flat spacetime limit, the Poincar\'e algebra is recovered in all these theories. As it is mentioned, for instance, in \cite{Bojowald2012,Bojowald2010,Brahma2018}, the Poincar\'e algebra can be obtained as a subalgebra of the off-shell algebra of constraints \eqref{gen_algebra} by considering the Minkowski spacetime (i.e., $N=1$, $N^i=0$ and $h_{ij}=\delta_{ij}$) and looking to the transformations generated by the Hamiltonian and momentum constraints associated with the linear functions
\begin{equation}
\alpha=\delta\mu^0+x_i\delta\omega^{i0},\qquad\qquad \alpha^i=\delta\mu^i+x_j\delta\omega^{ji}.\label{eq:genfun1}
\end{equation}
The algebra of constraints \eqref{gen_algebra} reads then
\begin{align}
&\left\{\mathcal{H}[\alpha],\mathcal{H}[\beta]\right\}=\mathcal{D}\left[\lambda F'(0)^2\delta^{ij}\left(\beta\partial_j\alpha-\alpha\partial_j\beta\right)\right],\nonumber\\
&\left\{\mathcal{D}[\alpha^i],\mathcal{H}[\alpha]\right\}=\mathcal{H}\left[\mathcal{L}_{\alpha^i}\alpha\right],\nonumber\\
&\left\{\mathcal{D}[\alpha^i],\mathcal{D}[\beta^j]\right\}=\mathcal{D}\left[\mathcal{L}_{\alpha^i} \beta^j\right].
\end{align}
We can now exploit the observation, made at the end of the previous section, that the absolute value of $\lambda$ can be changed arbitrarily in order to write the algebra [assuming $F'(0)\neq0$] in the form
\begin{align}
&\left\{\mathcal{H}[\alpha],\mathcal{H}[\beta]\right\}=\mathcal{D}\left[\theta\delta^{ij}\left(\beta\partial_j\alpha-\alpha\partial_j\beta\right)\right],\nonumber\\
&\left\{\mathcal{D}[\alpha^i],\mathcal{H}[\alpha]\right\}=\mathcal{H}\left[\mathcal{L}_{\alpha^i}\alpha\right],\nonumber\\
&\left\{\mathcal{D}[\alpha^i],\mathcal{D}[\beta^j]\right\}=\mathcal{D}\left[\mathcal{L}_{\alpha^i} \beta^j\right],\label{eq:hdefalg}
\end{align}
where $\theta=\pm1$. The value $\theta=1$ corresponds to general relativity, while $\theta=-1$ to Euclidean general relativity (see, e.g., \cite{Grain2016} and references therein). These equations can be directly used in order to extract the Poincar\'e algebra (for completeness, this is shown in App. \ref{sec:Poinc_calc}). This implies that all these theories contain either the  Poincar\'e algebra or its Euclidean counterpart.

\section{A class of Lagrangian theories \label{sec:sqrt}}

In the previous sections we have worked with a class of Hamiltonian theories that display an arbitrary function $F(x)$ of a particular combination of the variables in the phase space, but that still contain the same number of degrees of freedom as general relativity. We have illustrated that for all these theories [satisfying a non-degeneracy condition $F'(0)\neq0$], the Poincar\'e algebra is recovered in a suitable (standard) limit of the non-linear algebra of constraints. This is a first indication that even if these theories may seem different from general relativity, this may be an artifact of the way in which these are constructed, and at the end of the day these may turn out to be physically equivalent to the latter.

A clear possibility is that these theories could be obtained from the usual Hamiltonian formulation of general relativity by means of a non-linear field redefinition. However, this is fairly difficult to show in general, unless one is able to find by inspection the particular field redefinition that does the job (which, in any case, would not be a systematic approach). Hence, a different strategy is needed in order to learn more about these theories and their relation to general relativity. In particular, in this section we pick up a specific theory among the Hamiltonian theories previously studied, and switch to its Lagrangian formulation in order to illustrate a number of important features. The subsequent discussion is not limited to this particular theory but applies to a large class of Lagrangian theories, as detailed in Sec. \ref{sec:gen}. However, we will use this particular example in order to motivate the different steps in the analysis. This class of Lagrangian theories is not necessarily in one-to-one correspondence with the class of Hamiltonian theories discussed in Sec. \ref{sec:hamfam}, although we expect that the results below will be shared by all these Hamiltonian theories, as we have shown that the Poincar\'e algebra is recovered in all of them.

For a generic function $F(x)$, it is not easy to find the corresponding Lagrangian. However, one of the Lagrangian examples given in \cite{Lin2017} leads precisely to a Hamiltonian of the form given in Eq. \eqref{gen_ham}. These authors call this theory ``square root gravity'', which is defined by the Lagrangian density
\begin{equation}\label{sqrt_lag}
\mathcal{L}=\sqrt{h}N\left(M^4\sqrt{\left(c_1+c_2\mathcal{K}\right)\left(c_3+c_4R\right)}-c_5\right).
\end{equation}
Here, $M$ and $\{c_n\}_{n=1}^5$ are real constants with the appropriate dimensions, and $\mathcal{K}=K_{ij}K^{ij}-K^2$. It is worth mentioning that in the $c_1=c_3=c_5=0$ case, the Baierlein-Sharp-Wheeler action \cite{Baierlein} is recovered (see also \cite{Murchadha2012}).

The associated Hamiltonian is
\begin{equation}
\mathcal{H}=\sqrt{h}\left(-M^4\sqrt{c_1c_4}\sqrt{R-\frac{4}{c_2c_4}\frac{\Pi}{h}+\frac{c_3}{c_4}}+c_5\right).
\end{equation}
It is therefore clear that this theory belongs to the class introduced in Sec. \ref{sec:hamfam}, with $F(x)=-M^4\sqrt{c_1c_4}\sqrt{x+c_3/c_4}+c_5$ and $\lambda=-4/c_2c_4$. In particular, from our discussion above it follows that: (i) this theory contains two degrees of freedom (as discussed as well in \cite{Lin2017}), and (ii) if the constants are chosen such that Minkowski spacetime is a solution, these two degrees of freedom in $h_{ij}$ carry a representation of the Poincar\'e group. These two features (satisfied by all the Hamiltonian theories introduced before) suggest strongly that these two degrees of freedom must correspond to the usual two polarizations of the gravitational field in linearized general relativity. Let us show this explicitly in the following.

\subsection{Linearized field equations \label{sec:linear}}

Before linearizing the field equations, let us remark that not all the constants in Eq. \eqref{sqrt_lag} are independent. If we consider a constant rescaling of the lapse,
\begin{equation}
N\longrightarrow\zeta N,
\end{equation}
with $\zeta\in\mathbb{R}^+$, then Eq. \eqref{sqrt_lag} becomes
\begin{equation}\label{eq:3.4}
\mathcal{L}=\zeta\sqrt{h}N\left( M^4\sqrt{\left(c_1+c_2\mathcal{K}/\zeta^2\right)\left(c_3+c_4R\right)}- c_5\right).
\end{equation}
This Lagrangian is real when evaluated on a Minkowski spacetime if and only if $c_1c_3>0$. Moreover, it is always possible to choose $\zeta^2=|c_2c_3/c_1c_4|$ and rescale $M^4$ and $c_5$ accordingly so that  (\ref{eq:3.4}) becomes
\begin{equation}\label{sqrt_lag2}
\mathcal{L}=\sqrt{h}N\left(M^4\sqrt{\left(1+\theta c_4\mathcal{K}/c_3\right)\left(1+c_4R/c_3\right)}-c_5\right),
\end{equation}
with $\theta=\pm1$. Hence, we see that the coefficients in front of $\mathcal{K}$ and $R$ inside the square root can be chosen to be the same, up to a sign; this sign is related to the Euclidean or Lorentzian character of the theory, so we will choose in the following $\theta=1$ which corresponds to the Lorentzian sector.

Let us consider an expansion of the Lagrangian \eqref{sqrt_lag2} such that $N\longrightarrow 1+N$ and $h_{ij}\longrightarrow\delta_{ij}+h_{ij}$; alternatively,
\begin{equation}
g_{\mu\nu}=\eta_{\mu\nu}+h_{\mu\nu},
\end{equation}
where $h_{00}=N$ and $h_{0i}=\delta_{ij}N^j$. In practice, we can arrange the different orders in $h_{\mu\nu}$ in terms of the different orders in $\sqrt{h}N$ and the curvature scalars $\mathcal{K}$ and $R$. Let us choose $c_5$ requiring that there is no term proportional to $\sqrt{h}N$ only, namely $c_5=M^4$ (otherwise we would have to consider a non-zero cosmological constant, which however does not change the physical results). 

Renaming $c=c_4/c_3$ and taking into account that $\mathcal{K}$ is at least of second order in $h_{\mu\nu}$, while $R$ is of first order, Eq. \eqref{sqrt_lag2} can be expanded as
\begin{equation}
\mathcal{L}=cM^4\sqrt{-g}\left\{\left(\mathcal{K}+R\right)-\frac{c}{4}R^2\right\}+...
\end{equation}
where the dots indicate higher orders in $h_{\mu\nu}$ (that is, cubic and higher-order terms, which will be dealt with in Secs. \ref{sec:3pt} and \ref{sec:higher}, respectively).

The combination $\mathcal{K}+R$ can be recognized as the 4-dimensional Ricci scalar (up to a boundary term); hence this part of the Lagrangian leads to the Fierz-Pauli Lagrangian $\mathcal{L}_{\rm FP}$ at lowest order in $h_{\mu\nu}$. For the quadratic term in $R$, we just need to recall (e.g., \cite{Percacci2017}) that $R^{(0)}=0$ and $R^{(1)}=\partial_i\partial_jh^{ij}-\partial_i\partial^ih^j_{\,j}$, so that (the overall multiplicative constant is irrelevant)
\begin{equation}
\frac{1}{cM^4}\mathcal{L}=\mathcal{L}_{\rm FP}-\frac{c}{4}(\partial_i\partial_jh^{ij}-\partial_i\partial^ih^j_{\,j})^2.
\end{equation}
This action is invariant under the linear diffeomorphisms characteristic of Fierz-Pauli theory,
\begin{equation}\label{eq:lindiff}
h_{\mu\nu}\rightarrow h_{\mu\nu}+\partial_{(\mu}\xi_{\nu)}.
\end{equation}
On the other hand, the equations of motion are given by
\begin{align}\label{eq.11}
&\square h_{\mu\nu}-\partial_\mu\partial^\rho h_{\rho\nu}-\partial_\nu\partial^\rho h_{\rho\mu}+\eta_{\mu\nu}\partial_\rho\partial_\sigma h^{\rho\sigma}-\eta_{\mu\nu}\square h+\partial_\mu\partial_\nu h\nonumber\\
&-c\left(\partial_i\partial_j\partial_k\partial_lh^{kl}-\partial_i\partial_j\Delta \tilde{h}-\delta_{ij}\partial_k\partial_l\Delta h^{kl}+\delta_{ij}\Delta^2\tilde{h}\right)\delta^i_\mu\delta^j_\nu=0.
\end{align}
In this equation, we have defined $\square=\partial_\mu\partial^\mu$, $\Delta=\partial_i\partial^i$, and $\tilde{h}=h^j_{\,j}$. In particular, the $\mu=\nu=0$ component of the field equations is the same as in Fierz-Pauli theory, and takes the form
\begin{equation}
\square h_{00}+2(\partial_0)^2h_{00}+2\partial_0\partial^ih_{i0}+(\partial_0)^2h_{00}-2\partial_0\partial^ih_{i0}-\partial_i\partial_j h^{ij}+\square h+(\partial_0)^2 h=0,
\end{equation}
which can be arranged simply as
\begin{equation}\label{eq:00comp}
\Delta \tilde{h}-\partial_i\partial_j h^{ij}=0.
\end{equation}
This equation implies, as in Fierz-Pauli theory, that in the absence of matter it is possible to choose the so-called transverse and traceless gauge (e.g., \cite{Bishop2016}). Using the spatial diffeomorphisms, it is possible to choose a gauge in which $\partial_j h^{ij}=0$, but then Eq. \eqref{eq:00comp} reduces to $\Delta\tilde{h}=0$, which is the necessary and sufficient condition that allows the residual gauge transformations to gauge away the spatial trace $\tilde{h}$. Alternatively, the linearized field equations \eqref{eq.11} reduce to the Fierz-Pauli equations; to show this it is not even necessary to pick up a specific gauge, but rather to realize that Eq. \eqref{eq:00comp} directly implies that the term proportional to $c$ in Eq. \eqref{eq.11} vanishes, so that Eq. \eqref{eq.11} is written as
\begin{equation}
\mathcal{O}^{\mu\nu,\alpha\beta}h_{\alpha\beta}=0,\label{eq:fpeqs}
\end{equation}
where 
\begin{equation}
\mathcal{O}^{\mu\nu,\alpha\beta}=\eta^{\mu\left(\alpha\right.}\eta^{\left.\beta\right)\nu}\square-\eta^{\nu\left(\alpha\right.}\partial^{\left.\beta\right)}\partial^\mu-\eta^{\mu\left(\alpha\right.}\partial^{\left.\beta\right)}\partial^\nu+\eta^{\mu\nu}\partial^\alpha\partial^\beta-\eta^{\mu\nu}\eta^{\alpha\beta}\square-\eta^{\alpha\beta}\partial^\mu\partial^\nu.
\end{equation}

In summary, we have shown that the two degrees of freedom described by the theory with Lagrangian \eqref{sqrt_lag} correspond, in the linear limit, to the two usual polarizations of gravitons. This result is not accidental, as we knew in advance that linearizing the algebra of constraints of this theory leads to the Poincar\'e group. That the tensor field $h_{ij}$ transforms in the usual way under Poincar\'e transformations, combined with the fact that it encodes only two degrees of freedom, does not leave much wiggle room (the discussion above is not needed in order to realize this, but it is a nice explicit illustration of this general assertion).

\subsection{Propagator \label{sec:prop}}

Aside from the on-shell properties of the linearized field equations, the form of the propagator will be important in order to analyze the non-linear interactions. The propagator is an off-shell quantity, and therefore may show differences without necessarily implying the existence of physical differences. Let us start by writing the field equations \eqref{eq.11} in momentum space, and in the form of Eq. \eqref{eq:fpeqs}; i.e., in terms of the operator
\begin{align}\label{eq:difop1}
&\mathcal{O}^{\mu\nu,\alpha\beta}+c\delta^\mu_i\delta^\nu_j\delta^\alpha_k\delta^\beta_l\left(p^ip^jp^kp^l-\bm{p}^2p^ip^j\delta^{kl}-\bm{p}^2p^kp^l\delta^{ij}+\bm{p}^4\delta^{ij}\delta^{kl}\right)\nonumber\\
&=\mathcal{O}^{\mu\nu,\alpha\beta}+c\delta^\mu_i\delta^\nu_j\delta^\alpha_k\delta^\beta_l\left(p^ip^j-\bm{p}^2\delta^{ij}\right)\left(p^kp^l-\bm{p}^2\delta^{kl}\right)\nonumber\\
&=\mathcal{O}^{\mu\nu,\alpha\beta}+c\mathcal{A}^{\mu\nu}\mathcal{A}^{\alpha\beta},
\end{align}
where $\mathcal{O}^{\mu\nu,\alpha\beta}$ is the operator that appears in the field equations of Fierz-Pauli theory. The factorization property of the terms proportional to $c$ is quite important in what follows below. Let us recall that the field equations \eqref{eq.11} are invariant under linearized diffeomorphisms \eqref{eq:lindiff}. This implies that, as in Fierz-Pauli theory, we need to partially fix the gauge redundancy in order to be able to find the propagator. In order to keep the discussion as close as possible to the Fierz-Pauli case, let us fix the de Donder gauge $\partial_\mu h^{\mu\nu}=\partial^\nu h/2$ \cite{Hinterbichler2011,Ortin2015}. Let us recall that, in this gauge,
\begin{equation}\label{eq:dedonderop}
\mathcal{O}^{\mu\nu,\alpha\beta}=\frac{p^2}{2}(\eta^{\mu\alpha}\eta^{\nu\beta}+\eta^{\nu\alpha}\eta^{\mu\beta}-\eta^{\mu\nu}\eta^{\alpha\beta}).
\end{equation}
The inverse of this operator $\mathcal{F}_{\alpha\beta\rho\sigma}$, namely the quantity that verifies
\begin{equation}
\mathcal{O}^{\mu\nu,\alpha\beta}\mathcal{F}_{\alpha\beta\rho\sigma}=\frac{1}{2}(\delta^\mu_\rho\delta^\nu_\sigma+\delta^\mu_\sigma\delta^\nu_\rho),\label{eq:invnorm}
\end{equation}
is given by
\begin{equation}
\mathcal{F}_{\alpha\beta\rho\sigma}=\frac{1}{2p^2}(\eta_{\alpha\sigma}\eta_{\beta\rho}+\eta_{\alpha\rho}\eta_{\beta\sigma}-\eta_{\alpha\beta}\eta_{\sigma\rho}).
\end{equation}
The differential operator in the field equations \eqref{eq.11} has however a more complex expression, so its inverse has additional terms. We cannot use directly Eq. \eqref{eq:difop1}, as we need to impose the de Donder gauge, which in particular implies constraints between $h_{ij}$ and $h_{00}$. It is not difficult to show that the relevant expression to analyze is given by
\begin{equation}\label{eq:difop2}
\mathcal{O}^{\mu\nu,\alpha\beta}+c\mathcal{O}^{\mu\nu,00}\mathcal{O}^{00,\alpha\beta}.
\end{equation}
Note that the only difference with respect to Eq. \eqref{eq:difop1} is that the differential operator $\mathcal{A}^{\mu\nu}$ is identified with $\mathcal{O}^{00,\mu\nu}$ in the de Donder gauge, which can be obtained from Eq. \eqref{eq:dedonderop}. The problem of finding the inverse of the differential operator \eqref{eq:difop2} is a variant of the simpler calculation of the photon propagator in quantum electrodynamics (e.g., \cite{Zee2003}); for completeness, all  the necessary details are given in the following.

A detailed inspection of the problem suggests the ansatz
\begin{equation}
\mathcal{D}_{\alpha\beta\rho\sigma}=\mathcal{F}_{\alpha\beta\rho\sigma}+cB(p)\mathcal{W}_{\alpha\beta}\mathcal{W}_{\rho\sigma},
\end{equation}
where $B(p)$ is a function of $p^2$ and $\mathcal{W}_{\rho\sigma}$ is defined using Eq. \eqref{eq:invnorm}, namely
\begin{equation}\label{eq:rel1}
\mathcal{W}_{\rho\sigma}=p^2\mathcal{O}^{\alpha\beta,00}\mathcal{F}_{\alpha\beta\rho\sigma}=p^2\delta^0_\rho\delta^0_\sigma.
\end{equation}
This tensor satisfies
\begin{equation}\label{eq:rel2}
\mathcal{O}^{\mu\nu,\alpha\beta}\mathcal{W}_{\alpha\beta}=p^2\mathcal{O}^{\mu\nu,00}=p^2\mathcal{A}^{\mu\nu}.
\end{equation}
We can write therefore
\begin{align}
&[\mathcal{O}^{\mu\nu,\alpha\beta}+c\mathcal{O}^{\mu\nu,00}\mathcal{O}^{00,\alpha\beta}][\mathcal{F}_{\alpha\beta\rho\sigma}+cB(p)\mathcal{W}_{\alpha\beta}\mathcal{W}_{\rho\sigma}]\nonumber\\
&=\mathcal{O}^{\mu\nu,\alpha\beta}\mathcal{F}_{\alpha\beta\rho\sigma}+c\mathcal{O}^{\mu\nu,\alpha\beta}B(p)\mathcal{W}_{\alpha\beta}\mathcal{W}_{\rho\sigma}+c\mathcal{O}^{\mu\nu,00}\mathcal{O}^{00,\alpha\beta}\mathcal{F}_{\alpha\beta\rho\sigma}+c^2B(p)\mathcal{O}^{\mu\nu,00}\mathcal{O}^{00,\alpha\beta}\mathcal{W}_{\alpha\beta}\mathcal{W}_{\rho\sigma}\nonumber\\
&=\mathcal{O}^{\mu\nu,\alpha\beta}\mathcal{F}_{\alpha\beta\rho\sigma}+cp^2B(p)\mathcal{O}^{\mu\nu,00} \mathcal{W}_{\rho\sigma}+\frac{c}{p^2}\mathcal{O}^{\mu\nu,00}\mathcal{W}_{\rho\sigma}+c^2B(p)\mathcal{O}^{\mu\nu,00}\mathcal{O}^{00,\alpha\beta}\mathcal{W}_{\alpha\beta}\mathcal{W}_{\rho\sigma}.
\end{align}
In order to write the last identity we have used Eqs. \eqref{eq:rel1} and \eqref{eq:rel2}. Imposing the equivalent of Eq. \eqref{eq:invnorm}, we can solve for $B(p)$ as
\begin{equation}
B(p)=-\frac{1}{p^2(p^2+c\mathcal{O}^{00,\alpha\beta}\mathcal{W}_{\alpha\beta})}.
\end{equation}

We just have to evaluate
\begin{equation}
\mathcal{O}^{00,\alpha\beta}\mathcal{W}_{\alpha\beta}=p^2\mathcal{O}^{00,00}=\frac{p^4}{2},
\end{equation}
in order to obtain the full propagator
\begin{equation}\label{eq:fullprop}
\mathcal{D}_{\alpha\beta\rho\sigma}=\mathcal{F}_{\alpha\beta\rho\sigma}-\frac{c}{1+cp^2/2}\delta^0_\alpha\delta^0_\beta\delta^0_\rho\delta^0_\sigma.
\end{equation}
We see that the propagator is not the same as in Fierz-Pauli theory, which is reasonable as off-shell equivalence is not guaranteed (nor needed in order to show physical equivalence). However, the propagator is still a function of the covariant combination of $p^2$ only and, for large $p^2$, its behavior is the same as in Fierz-Pauli theory, namely $1/p^2$. These features will be of importance later, in Sec. \ref{sec:higher}.

\vspace{0.5cm}
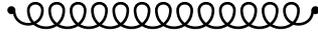
\begin{figure}[h]
\begin{center}
\begin{fmffile}{prop}
\begin{fmfgraph*}(40,1)
\fmfleft{i1}
\fmfright{o1}
\fmf{gluon}{i1,o1}
\fmfv{decor.shape=circle,decor.filled=full,decor.size=1thick}{i1}
\fmfv{decor.shape=circle,decor.filled=full,decor.size=1thick}{o1}
\end{fmfgraph*}
\end{fmffile}
\end{center}
\caption{The propagator is a function of $p^2$ with a leading dependence $1/p^2$.}
\label{fig:prop}
\end{figure}
\subsection{On-shell 3-point amplitudes \label{sec:3pt}}

Let us now consider the next order in the perturbative expansion of Eq. \eqref{sqrt_lag2}:
\begin{equation}
\mathcal{L}=cM^4\sqrt{-g}\left\{\left(\mathcal{K}+R\right)+cR^2+c\mathcal{K}R+c^2R^3\right\}+...
\end{equation}
where again we are omitting terms, which this time are at least of fourth order in $h_{\mu\nu}$. The first piece, proportional to $\sqrt{-g}(\mathcal{K}+R)$, leads to the usual 3-point amplitudes. On the other hand, the additional terms contain the new interaction vertices
\begin{align}
&\sqrt{-g}\,R^2=[R^{(1)}]^2-\frac{1}{2}h[R^{(1)}]^2+2 R^{(1)}R^{(2)}+...\nonumber\\
&\sqrt{-g}\,\mathcal{K}R=\mathcal{K}^{(2)}R^{(1)}+...\nonumber\\
&\sqrt{-g}\,R^3=[R^{(1)}]^3+...\label{eq:offsh3}
\end{align}
But we have shown that Eq. \eqref{eq:00comp} is satisfied at the linear level, which implies
\begin{equation}
R^{(1)}=\partial_i\partial_j h^{ij}-\Delta \tilde{h}=0.\label{eq:ric3pt}
\end{equation}
Note that this quantity above is precisely the $\mu=\nu=0$ component of the field equations \eqref{eq:fpeqs}, $\mathcal{O}^{00,\alpha\beta}h_{\alpha\beta}=0$, up to a sign. Therefore, if just one of the legs of the 3-point amplitude is on-shell, the contributions coming from the $[R^{(1)}]^3$ vertex vanish identically; no differences from general relativity can result from these vertices. Moreover, noting that $R^{(1)}=-\mathcal{O}^{00,\alpha\beta}h_{\alpha\beta}$ appears in all the 3-point vertices in Eq. \eqref{eq:offsh3}, if follows that this observation extends to all these vertices when all the external particles are on-shell (i.e., when the amplitude is on-shell).

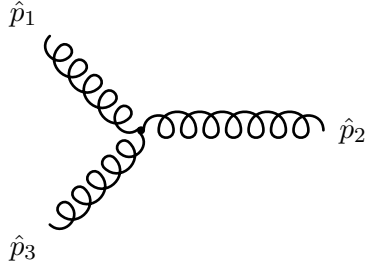
\begin{figure}[h]
\begin{center}
\begin{fmffile}{3pt1}
\begin{fmfgraph*}(40,25)
\fmfleft{i1,i2}
\fmfright{o1}
\fmflabel{$\hat{p}_1$}{i2}
\fmflabel{$\hat{p}_2$}{o1}
\fmflabel{$\hat{p}_3$}{i1}
\fmf{gluon}{i2,v1}
\fmf{gluon}{i1,v1}
\fmf{gluon}{o1,v1}
\fmfv{decor.shape=circle,decor.filled=full,decor.size=1thick}{v1}
\end{fmfgraph*}
\end{fmffile}
\end{center}
\caption{The on-shell 3-point amplitudes (of complex momenta) are the same as in general relativity. As discussed in the next section, these are the building blocks for constructing on-shell $n$-point amplitudes of real momenta for $n\geq4$, which makes this equivalence fundamental for our discussion.}
\label{fig:3pt}
\end{figure}

Let us recall that 3-point amplitudes of massless particles become trivial on-shell \cite{Elvang2015,Elvang2013}. However, if the momenta of the external particles are considered to be complex, this is no longer the case \cite{Witten2003} (see also \cite{Benincasa2007b}). This allows to write expressions for the 3-point on-shell amplitudes which are non-zero for complex momenta, but become trivial for real momenta. The argument above following which the contributions from the non-covariant interaction vertices (e.g., $[R^{(1)}]^3$) vanish if the external legs are on-shell, applies as well to complex momenta. Hence, we can conclude that the on-shell 3-point amplitudes (of complex momenta) for the theory with Lagrangian density \eqref{sqrt_lag2} are the same as in general relativity.

\subsection{Brief pause: most general theories compatible with these results \label{sec:gen}}

The results in Secs. \ref{sec:linear}, \ref{sec:prop} and \ref{sec:3pt} are not particular to the theory with Lagrangian density \eqref{sqrt_lag}. We have made this observation above, but we would like to show it more explicitly at this point in the discussion. Let us consider the Lagrangian density
\begin{equation}\label{eq.20}
\mathcal{L}=\sqrt{-g}\,G(\mathcal{K},R),
\end{equation}
where $G(\mathcal{K},R)$ is a function of $\mathcal{K}$ and $R$, satisfying the conditions derived in \cite{Lin2017} so that the theory contains two local degrees of freedom. It is interesting to note that Eq. \eqref{eq.20} is general enough to include all the explicit examples constructed in \cite{Lin2017}. This function can be expanded around Minkowski spacetime as
\begin{equation}
\left.\frac{\partial \mathcal{L}}{\partial R}\right|_{R=\mathcal{K}=0}R+\left.\frac{\partial \mathcal{L}}{\partial \mathcal{K}}\right|_{R=\mathcal{K}=0}\mathcal{K}+\left.\frac{1}{2}\frac{\partial^2 \mathcal{L}}{\partial \mathcal{K}\partial R}\right|_{R=\mathcal{K}=0}R\mathcal{K} +\left.\frac{1}{2}\frac{\partial^2 \mathcal{L}}{\partial R^2}\right|_{R=\mathcal{K}=0}R^2+\left.\frac{1}{6}\frac{\partial^3 \mathcal{L}}{\partial R^3}\right|_{R=\mathcal{K}=0}R^3+...
\end{equation}
Unless $\left.\partial \mathcal{L}/\partial R\right|_{R=\mathcal{K}=0}=0$ or $\left.\partial \mathcal{L}/\partial \mathcal{K}\right|_{R=\mathcal{K}=0}=0$, it is always possible to perform field redefinitions (constant rescalings) as we have done previously in order to reconstruct the 4-dimensional Ricci scalar up to an irrelevant boundary term. We have then an expansion in $\mathcal{K}$ and $R$ that is precisely of the form that has been used in Secs. \ref{sec:linear}, \ref{sec:prop} and \ref{sec:3pt}, so that the results of these sections equally apply to all the theories with Lagrangian \eqref{eq.20} (implying, in particular, the equivalence of the linearized field equations and 3-point amplitudes with that of general relativity). Let us note that $\left.\partial \mathcal{L}/\partial \mathcal{K}\right|_{R=\mathcal{K}=0}$ must be non-zero in order to have a kinetic term for the field $h_{\mu\nu}$, so the only possibility to avoid this general result within this family is considering theories in which $\left.\partial \mathcal{L}/\partial R\right|_{R=\mathcal{K}=0}=0$. We analyze this case separately in Sec. \ref{sec:nogravs}, showing that this kind of theory does not even contain gravitons at the linearized level, which raises doubts about whether or not these describe gravitational theories (and is also disastrous from an observational perspective).

\subsection{Higher orders in the perturbative expansion \label{sec:higher}}

Up to now, we have shown that the theory defined by the Lagrangian density in Eq. \eqref{sqrt_lag} is equivalent to general relativity for its two lowest orders: at the linear level the field equations are equivalent to that of Fierz-Pauli theory, and at first order in the interactions, the 3-point amplitudes are the same as in general relativity. Moreover, we have discussed that the latter property is an unavoidable consequence of the former in all theories of the fairly general form \eqref{eq.20}. Armed with this knowledge, we can use now a battery of results from on-shell scattering amplitudes in order to obtain non-trivial information about the higher orders in the perturbative expansion. The goal of this section is simple: we know that theories of the form \eqref{eq.20} that satisfy $\left.\partial \mathcal{L}/\partial R\right|_{R=\mathcal{K}=0}\neq0$ and $\left.\partial \mathcal{L}/\partial \mathcal{K}\right|_{R=\mathcal{K}=0}\neq0$ have the same polarizations of the gravitational field at the linear level and the same 3-point amplitudes as general relativity; in this section we study whether or not this equivalence holds for $n$-point amplitudes with $n\geq 4$.

In order to achive our goal, we will have to take a detour from our previous discussion in terms of Hamiltonian and Lagrangian mechanics and consider a language which is more powerful in order to deal with the perturbative structure of non-linear theories. This language is that of scattering amplitudes in quantum field theory, in its modern incarnation using spinor-helicity variables which are specially suited for the description of massless particles. In order to avoid introducing the notation we will be using in this section, we will just stick to the notation conventions in \cite{Elvang2015} (see also \cite{Elvang2013}) for the spinor-helicity variables and the polarization tensors of the gravitational field. Some of these ingredients are briefly reviewed in App. \ref{sec:app2} in order to make the paper reasonably self-contained.

We have studied on-shell 3-point amplitudes in Sec. \ref{sec:3pt}. However, for the purposes of this section, we will need to take into account some properties of the off-shell 3-point vertices of the actions with Lagrangian density \eqref{eq.20}. The most important observation is that, while in general relativity any interaction vertex only contains two powers of momenta, the theories \eqref{eq.20} do not verify this requirement. In particular, from the last line in Eq. \eqref{eq:offsh3} we can read that off-shell 3-point vertices can have up to 6 powers of the (spatial) momenta. This feature forbids applying previous uniqueness results \cite{Cachazo2005,Benincasa2007,Hall2008,ArkaniHamed2008} that use an extension of the Britto-Cachazo-Feng-Witten (BCFW) relations \cite{Britto2004,Britto2005} but assume that interaction vertices are at most quadratic in the momenta (that higher-derivative interaction vertices are problematic is well-known \cite{Cohen2010}). Technically, this follows from the necessity of showing that contributions from Feynman diagrams remain bounded for large (complex) momenta; that vertices may contain higher powers of momenta spoil the bounds that can be obtained for the quadratic case. The most virulent behavior comes from off-shell 3-point vertices with 6 powers of the momenta, namely the ones in the last line of Eq. \eqref{eq:offsh3}. However, these contributions can only come from internal vertices, as we have shown in Sec. \ref{sec:3pt} that these vertices vanish when at least one of the legs stitched to it is on-shell.

Let us start with the $n$-point amplitude of $n$ gravitons with the same helicity,\linebreak $A_n(1^+2^+...\,n^+)$, or just with the helicity of one graviton flipped, $A_n(1^-2^+...\,n^+)$. In general relativity these amplitudes vanish for any value of $n$. There is a simple argument to show that this must be the case by counting the powers of momenta in Feynman diagrams (see, e.g., Sec. 2.7 in \cite{Elvang2015}). The argument boils down to the observation that this amplitude can be non-zero only if vertices contribute at least with $2n$ powers of the momenta that can be contracted with the $2n$ polarization vectors \eqref{eq:polvec+} and \eqref{eq:polvec-} of external gravitons [no powers of the momenta can be provided by propagators, which is true both in general relativity and in the theories analyzed here; recall Eq. \eqref{eq:fullprop}]. If this is not the case, at least two of these polarization vectors will be contracted with each other, and this combination can be shown to be zero up to a gauge transformation, implying that the overall amplitude vanishes. While it is impossible to obtain these $2n$ powers of the momenta in the case of general relativity (and, in general, for any theory in which interaction vertices are quadratic in the momenta), here we have to be more careful with the interaction vertices containing higher powers of momenta. However, the potentially most virulent behavior, coming from the $[R^{(1)}]^3$ off-shell 3-point vertex in Eq. \eqref{eq:offsh3}, turns out to be innocuous as the corresponding 6 powers of the momenta cannot be coupled to the polarization vectors of external particles. Taking a look at Eq. \eqref{eq:ric3pt}, we see that the first part can be gauged away (off-shell), while the second one contributes with a factor $p_i p^i$ that has no free indices to be contracted with the indices of polarization vectors. Alternatively, in the de Donder gauge, $R^{(1)}$ is simply proportional to the scalar $p^2$. The remaining 3-point vertices $\mathcal{K}^{(2)}R^{(1)}$ and $R^{(2)}R^{(1)}$ only have 2 powers of the momenta that can be contracted with the $\epsilon^\mu_+$, which leads to the same maximum number of momenta as in general relativity. Let us therefore turn our attention to 4-point vertices, from which $[\mathcal{K}^{(2)}]^2$ and $[R^{(2)}]^2$ display the worst behavior, with a $p^4$ dependence. However, counting the number of vertices in the corresponding Feynman diagrams (see App. \ref{sec:app2}) shows that in this case the maximum possible number of momenta is $2n-4<2n$. Therefore, we conclude that
\begin{equation}
A_n(1^+2^+...\,n^+)=A_n(1^-2^+...\,n^+)=0,\qquad \forall n\geq3.\label{eq:0amp}
\end{equation}
This is a quite non-trivial statement that is characteristic of general relativity as well as Yang-Mills theories. This implies that, in order to have a non-zero amplitude, we need to flip the helicity of at least two gravitons (which leads to the so-called maximally helicity violating --MHV-- amplitudes \cite{BjerrumBohr2005}).

In order to deal with the MHV amplitudes $A_n(1^-2^-3^+...\,n^+)$ and other possibly non-vanishing amplitudes, we need to resort to more sophisticated arguments. As reviewed briefly in App. \ref{sec:app2}, the most powerful argument to deal with these amplitudes involves complexifying the external momenta in order to justify constructing $(n+1)$-point amplitudes from $n$-point amplitudes \cite{Elvang2015}; the BCFW recursion relation \cite{Britto2004,Britto2005} is a particular example of this procedure. This procedure permits to construct $A_n(1^-2^-3^+...\,n^+)$ as a sum of quadratic products of on-shell $(n-1)$-amplitudes evaluated on complex momenta. This recursive procedure is valid as long as certain technical conditions are satisfied by the complexification of the $n$-point amplitude $A_n(1^-2^-3^+...\,n^+)$. In particular, we will consider the particular complex extension of external momenta that was considered in the App. A of \cite{Benincasa2007} (see also \cite{Hall2008}) in order to obtain what these authors call the ``auxiliary'' recursion relations. This particular complex extension represents the optimal choice in general relativity, and therefore it will be also the optimal choice here, as the only change in the present discussion is the behavior of interaction vertices with the momenta. Note that we cannot apply directly the conclusions of these authors, as the theories we are interested in have interaction vertices which are not quadratic in the momenta.

Without loss of generality, let us consider an amplitude in which $N\geq n/2$ external legs have positive helicity; the case in which there is a larger number of external legs with negative helicity is completely parallel. If $s\in[1,N]$ labels all the external legs with positive helicity, and $t$ marks a single external leg with negative helicity, we define the complex extension
\begin{equation}
\hat{p}_s=p_s+zq_s,\qquad\qquad \hat{p}_t=p_t-z\sum_{s=1}^Nq_s,
\end{equation}
where $z\in\mathbb{C}$ and $\{q_s\}_{s=1}^N$ are complex vectors verifying certain requirements; specific expressions for these vectors are given in App. \ref{sec:app2}.

To show the validity of the recursion relations, let us exploit the fact that it is a sufficient condition that all Feynman diagrams contributing to a given amplitude vanish asymptotically for large $|z|$. Taking into account that Feynman diagrams are just the multiplication of diverse elements (external legs, interaction vertices and propagators), we just need to understand how these different elements scale with $|z|$:
\begin{itemize}
\item{$N+1$ external particles are shifted (that is, their momenta are complexified) such that their polarization tensors constructed from Eqs. \eqref{eq:polvec+} and \eqref{eq:polvec-} give each a leading contribution $1/|z|^2$.}
\item{All propagators in Eq. \eqref{eq:fullprop} are shifted (this property that depends only on the topology of the relevant Feynman diagrams was shown in \cite{Benincasa2007}) and therefore contribute with a $1/|z|$ factor each.}
\item{Vertices behave as $|z|^k$. Internal (i.e., off-shell) 3-point vertices display 6 powers of the spatial momenta instead of being just quadratic in the covariant momenta, which suggests a $|z|^{6}$ dependence. However, a careful inspection shows (as discussed in Secs. \ref{sec:prop} and \ref{sec:3pt}) that these off-shell vertices are proportional in the de Donder gauge to the covariant combination $(p^2)^3$, which means that these also scale as $|z|^3$. A similar comment applies to internal vertices that contain $R^{(1)}$.}
\end{itemize}
Putting all these elements together, for Feynman diagrams involving only $k$-point vertices the leading $z$ dependence goes like
\begin{equation}
A_n(z)\propto \left(z^{-2}\right)^{N+1}\left(z^{-1}\right)^{p}\left(z^k\right)^{v},
\end{equation}
where the first multiplicative factor is the contribution from the polarization tensors of the shifted external legs, $p$ is the number of propagators and $v$ is the number of vertices. For $k$-point vertices one has (App. \ref{sec:app2})
\begin{equation}
p=\frac{n-k}{k-2},\qquad\qquad v=\frac{n-2}{k-2}.
\end{equation}
The recursion relations can be valid only if $A_n(z)\propto z^w$ with $w<0$ (which in general relativity is always satisfied \cite{Benincasa2007}). This leads to the inequality
\begin{equation}\label{eq:fcond}
N>\frac{(k-1)n-(3k-4)}{2(k-2)}.
\end{equation}
It is straightforward to show that the most stringent of these conditions is obtained for $k=3$, which implies that interaction vertices with $k\geq4$ only give subleading contributions to the asymptotic behavior on $|z|$ of Feynman diagrams.

Overall, this leads to the main results of this section, together with Eq. \eqref{eq:0amp}:
\begin{itemize}
\item{$n=4$ and $n=5$: all amplitudes $A_n(1^{h_1}2^{h_2}...\,n^{h_n})$ can be constructed from 3-point amplitudes and are therefore the same as in general relativity.}
\item{$n\geq6$: the MHV amplitude $A_6(1^-2^-3^+4^+5^+6^+)$ is the same as in general relativity. This argument does not fix the form of $A_6(1^-2^-3^-4^+5^+6^+)$ or any of the remaining $n$-point amplitudes with $n>6$ that are not already contained in Eq. (\ref{eq:0amp}).}
\end{itemize}
In summary, we have applied state-of-the-art techniques of on-shell scattering amplitudes in order to calculate a sequence of scattering amplitudes which would be otherwise extremely difficult (and time-consuming) to calculate. This illustrates that on-shell techniques can be used in order to understand the physical content of modified gravity theories. The information extracted in this way (see Fig. \ref{fig:4p}) is quite interesting, as it uncovers unexpected cancellations of the additional (non-relativistic) interaction vertices, and strengthens the possibility that these theories are equivalent to general relativity. In particular, we have determined that differences might arise only for $6$-point amplitudes [in the $A_6(1^-2^-3^-4^+5^+6^+)$ amplitude] or higher.

Let us stress that this does not necessarily imply that it is likely that there will be differences. It is ubiquitous in the study of on-shell scattering amplitudes that unexpected cancelations occur, so that terms that may seemingly give additional contributions leave no trace. In fact, we already know that these kinds of cancellations indeed happen for the theories analyzed here. For instance, in the amplitudes $A_4(1^{h_1}2^{h_2}3^{h_3}4^{h_4})$ the contributions coming from the additional (i.e., not present in general relativity) 4-point vertices $[\mathcal{K}^{(2)}]^2$ and $[R^{(2)}]^2$ cancel identically. A similar observation follows for arbitrary $A_n(1^{h_1}2^{h_2}...n^{h_n})$ for $n\leq5$, and applies to all the possible interaction vertices that are involved in these amplitudes, including the most virulent $[R^{(1)}]^3$. This even applies to $A_n(1^+2^+...\,n^+)$ and $A_n(1^-2^+...\,n^+)$ for arbitrary values of $n$. It is not unreasonable to think that these cancellations will keep taking place for the remaining tree-level amplitudes, but that the on-shell arguments used here are not powerful enough to show this. Reaching a definitive conclusion is not possible at the moment; it may be that some generalization of these arguments will be successful, or additional calculations might uncover a counterexample. But we certainly think this is an interesting question that deserves further study.

\begin{figure}[H]
\begin{center}
\begin{equation}
\begin{gathered}
\begin{fmffile}{4pt1}
\begin{fmfgraph*}(40,25)
\fmfleft{i1,i2}
\fmfright{o1,o2}
\fmf{gluon}{i1,v1}
\fmf{gluon}{i2,v1}
\fmf{gluon}{v1,o1}
\fmf{gluon}{v1,o2}
\fmfv{decor.shape=circle,decor.filled=shaded,decor.size=.30w}{v1}
\end{fmfgraph*}
\end{fmffile}
\end{gathered}
=
\begin{gathered}
\begin{fmffile}{4pt2}
\begin{fmfgraph*}(40,25)
\fmfleft{i1,i2}
\fmfright{o2,o1}
\fmf{gluon}{i2,v1}
\fmf{gluon}{v1,i1}
\fmf{gluon}{v1,v2}
\fmf{gluon}{o1,v2}
\fmf{gluon}{v2,o2}
\fmfv{decor.shape=circle,decor.filled=full,decor.size=1thick}{v1}
\fmfv{decor.shape=circle,decor.filled=full,decor.size=1thick}{v2}
\end{fmfgraph*}
\end{fmffile}
\end{gathered}
+
\begin{gathered}
\begin{fmffile}{4pt3}
\begin{fmfgraph*}(40,25)
\fmfleft{i1,i2}
\fmfright{o1,o2}
\fmf{gluon}{i1,v1}
\fmf{gluon}{i2,v1}
\fmf{gluon}{v1,o1}
\fmf{gluon}{v1,o2}
\fmfv{decor.shape=circle,decor.filled=full,decor.size=1thick}{v1}
\end{fmfgraph*}
\end{fmffile}
\end{gathered}
\end{equation}
\end{center}
\caption{One of the results in this section is that the $A_4(1^{h_1}2^{h_2}3^{h_3}4^{h_4})$ is constructible from on-shell 3-point amplitudes of complex momenta using recursion relations, which in particular implies equivalence to general relativity at this particular order in the perturbative expansion. As in general relativity, this means that 4-point vertices carry no physical information, but are present in order to guarantee off-shell gauge invariance. This result is remarkable as it extends this observation from the 4-point vertices $\mathcal{K}^{(4)}$ and $R^{(4)}$ of general relativity, to include also $[\mathcal{K}^{(2)}]^2$ and $[R^{(2)}]^2$. Moreover, we have shown that perturbative equivalence with general relativity holds for $A_n(1^{h_1}2^{h_2}...4^{h_4})$ up to $n=5$ no matter the helicity configuration, and for selected configurations when $n>5$.}
\label{fig:4p}
\end{figure}

Before finishing this section, let us stress that a central point behind the rationale of using these on-shell methods is that we already know in advance that there are theories of the form \eqref{eq.20} that contain two local degrees of freedom, for instance the square root gravity with Lagrangian density \eqref{sqrt_lag}. This observation is even more important due to the fact that these theories generally contain higher-derivative operators, which is typically associated with the occurrence of additional degrees of freedom unless certain conditions are met \cite{Woodard2006,Klein2016,Crisostomi2017}. Without this information, even if one was able to show that the $n$-point amplitudes of the two graviton polarizations are the same as in general relativity, it would not be possible to discard the existence of additional degrees of freedom that may enter at higher orders in the perturbative expansion (i.e., the fact that only two local degrees of freedom are found may be due to considering a perturbative expansion around Minkowski spacetime \cite{Belenchia2016}; an example is given by the theory $\mathcal{L}=\sqrt{-g}(\mathcal{R}+\mathcal{R}^3)$ where $\mathcal{R}$ is the 4-dimensional Ricci scalar). This comment illustrates how intertwined the Hamiltonian analysis in Sec. \ref{sec:hamfam} and the discussion in this section are.

\section{Genuinely different theories}\label{sec:diff_theor}

In the previous sections we have used different probes of large classes of theories and found no evidence of differences with respect to general relativity. It would be interesting to find theories that still satisfy the conditions in \cite{Lin2017} ensuring that the number of degrees of freedom is the same as general relativity, but that are manifestly different from general relativity. In this section we analyze two kinds of theories that fall within this category, discussing their main properties and, in particular, their pathologies.

\subsection{Theories with no gravitons \label{sec:nogravs}}

As already noticed in \cite{Lin2017}, theories that do not contain the 3-dimensional Ricci tensor $R_{ij}$ in the Lagrangian density describe automatically two degrees of freedom . In particular, for theories of the form \eqref{eq.20} this would imply $\partial \mathcal{L}/\partial R=0$. The simplest example of this kind of theory is given by
\begin{equation}
\mathcal{L}=\sqrt{h}N\,\mathcal{K}.
\end{equation}
Let us now show that this theory is clearly inequivalent to general relativity, by looking at its Hamiltonian formulation: the momenta are given by
\begin{equation}
\pi^{ij}=2\left(K^{ij}-Kh^{ij}\right)\frac{1}{2}\sqrt{h},\qquad\pi_N=0,\qquad\pi_i=0,
\end{equation}
and the total Hamiltonian takes the usual form
\begin{equation}
H=
\int d^3x\,(N\mathcal{H}+N^i\mathcal{D}_i+\lambda_N\pi_N+ \lambda^i\pi_i),
\end{equation}
but with
\begin{equation}
\mathcal{H}=\frac{\Pi}{\sqrt{h}}.
\end{equation}
The secondary constraints are again $\mathcal{H}=0$ and $\mathcal{D}_i=0$. The main difference lies now in the algebra of constraints, which takes the form
\begin{align}\label{gen_algebra_nogravs}
&\left\{\mathcal{H}[\alpha],\mathcal{H}[\beta]\right\}=0,\nonumber\\
&\left\{\mathcal{D}[\alpha^i],\mathcal{H}[\alpha]\right\}=\mathcal{H}\left[\mathcal{L}_{\alpha^i}\alpha\right],\nonumber\\
&\left\{\mathcal{D}[\alpha^i],\mathcal{D}[\beta^j]\right\}=\mathcal{D}\left[\mathcal{L}_{\alpha^i} \beta^j\right].
\end{align}
Note the difference with respect to Eq. \eqref{gen_algebra}: the first bracket is identically vanishing. This has clear physical implications, which can be noticed for instance recalling our discussion in Sec. \ref{sec:poincare} and App. \ref{sec:Poinc_calc}: even if the theory describes two degrees of freedom, at linear level these excitations do not carry a representation of the Poincar\'e group (not even in some kind of low-energy limit). For instance, in the theory analyzed in this section, time translations and boosts commute, as well as two boosts (in the Poincar\'e group, these commutators will be proportional to spatial translations and rotations, respectively). Therefore, these theories are escaping equivalence with general relativity, but at the price of modifying drastically even the linear properties of the theory (i.e., not describing gravitons at the linear level). 

In particular, it is not clear if it is legitimate to call these theories ``gravitational'' theories \cite{Lin2017}. The linear excitations in this theory do not satisfy a wave equation, which would imply that there are no gravitational waves and therefore contradicts observational facts \cite{Abbott2016}. This is a reasonable outcome, which shows that tampering with the very properties of the carrier of the gravitational force generally has drastic physical consequences.

\subsection{Theories with a modified dispersion relation for gravitons \label{sec:moddisp}}

The previous example is interesting from a theoretical perspective but is, in some sense, too trivial. In this section we consider a class of examples that are more interesting physically.
In some sense these examples are complementary to the ones in the previous section: instead of considering a Lagrangian density that does not depend on $R_{ij}$, let us consider Lagrangian densities that are obtained by adding a piece that is independent from $K_{ij}$ (and also the lapse $N$). As shown in a brief section at the end of the paper \cite{Lin2017}, adding this term renders the Hamiltonian constraint second class, but it also introduces a tertiary constraint. Therefore, the number of degrees of freedom does not change even if the way the counting is performed is changed. This class of examples was also found in \cite{Comelli2014}. Phenomenological consequences for some of these theories (and extensions) are studied in \cite{Lin2018}. Here, we focus on the effects that these kind of terms have when added to the Einstein-Hilbert action; our conclusions stated below do not depend on this specific choice of starting point. For concreteness, let us consider
\begin{equation}
\mathcal{L}=\sqrt{h}N(\mathcal{K}+R)+\ell^2\sqrt{h}\,R_{ij}R^{ij},\label{eq:ric2pot}
\end{equation}
where $\ell$ is a constant with dimensions of length. \\ Let us now obtain the form of the field equations. The first piece in Eq. \eqref{eq:ric2pot} leads to the usual Fierz-Pauli equations, so we will focus on the second piece, for which we need to recall (e.g., \cite{Percacci2017}) that $R^{(1)}_{ij}=(\Delta h_{ij}-\partial_i\partial^kh_{kj}-\partial_j\partial^kh_{ki}+\partial_i\partial_j\tilde{h})/2$, so that the corresponding contribution to the field equations is proportional to
\begin{equation}\label{eq:lvterms}
\frac{1}{2}\left(\delta_i^{k}\delta_j^{l}\Delta-2\delta_i^{k}\partial_j\partial^l+\delta_{ij}\partial^k\partial^l\right)\left(\Delta h_{kl}-\partial_k\partial^mh_{ml}-\partial_l\partial^mh_{mk}+\partial_k\partial_l\tilde{h}\right)\delta^i_\mu\delta^j_\nu.
\end{equation}
We can see that the additional term describe higher-derivative corrections to the purely spatial components of the Fierz-Pauli equations, similarly to what happened in the square root gravity, Eq. \eqref{eq.11}. However, in the square root gravity these contributions were identically vanishing. Here we will see that the same does not apply.

First of all, let us recall that, as in the square root gravity, the $\mu=\nu=0$ component of the field equations is given by Eq. \eqref{eq:00comp}. This, in turn, implies that the trace of Eq. \eqref{eq:lvterms} vanishes on-shell: it is given by
\begin{align}
&\frac{1}{2}\left(\delta^{kl}\Delta+\partial^k\partial^l\right)\left(\Delta h_{kl}-\partial_k\partial^mh_{ml}-\partial_l\partial^mh_{mk}+\partial_k\partial_l\tilde{h}\right)\nonumber\\
&=\frac{3}{2}\Delta(\Delta \tilde{h}-\partial_i\partial_jh^{ij})=0.
\end{align}
We can therefore:
\begin{itemize}
\item{Impose the de Donder gauge $\partial^\mu \bar{h}_{\mu\nu}=0$, where $\bar{h}_{\mu\nu}=h_{\mu\nu}-\eta_{\mu\nu}h/2$.}
\item{Exploit the $\mu=0$, $\nu=0$ and $\mu=0$, $\nu=i$ components of the field equations to gauge away the $\bar{h}_{00}$ and $\bar{h}_{0i}$ components of the gravitational field.}
\item{Use the trace of the field equations, $\square \bar{h}=0$, to fix completely the residual gauge freedom setting $\bar{h}=0$.}
\end{itemize}
This implies that the field equations can be written simply as
\begin{equation}
\square h_{ij}+\ell^2\Delta^2 h_{ij}=0.
\end{equation}
This is a quite interesting equation that implies that this theory describes gravitons (with two degrees of freedom), but with modified dispersion relations. At least at low energies (measured in terms of the scale $\ell$), the usual Lorentz-invariant picture is recovered. This suggests that these theories are interesting alternatives to general relativity that reduce to the latter at low energies.

However, there is a feature of the theories which include ``potential'' terms in the 3-dimensional Ricci tensor $R_{ij}$ that must be kept in mind. The inclusion of this kind of potential term independent of the lapse $N$ and the extrinsic curvature $K_{ij}$ changes drastically the nature of the constraints of the theory. First of all, the form of the Hamiltonian and diffeomorphism constraints is unchanged. However, the Hamiltonian constraint is not automatically preserved by evolution, and an additional constraint must be added in order to ensure its preservation. This, in turn, implies that the Hamiltonian constraint is second class. Hence, the only remaining gauge symmetries are spatial diffeomorphisms. But here is where the trouble lies: instead of the very combination
\begin{equation}
\mathcal{K}=K_{ij}K^{ij}-K^2,
\end{equation}
that appears in the Lagrangian density \eqref{eq:ric2pot} and is characteristic of general relativity, spatial diffeomorphisms can only select the combination
\begin{equation}
K_{ij}K^{ij}-\mu K^2,
\end{equation}
with $\mu$ an arbitrary constant; this is for instance a well-known observation in the framework of Ho\v{r}ava-Lifshitz gravity \cite{Horava2009}. Equally well-known in this framework (e.g., \cite{Visser2011}) is the fact that a theory with a Lagrangian density linear in the combination just above can only contain two degrees of freedom if and only if $\mu=1$; for different values of this parameter, an additional degree of freedom (a scalar graviton) appears. Let us stress that the same result follows from our analysis in Sec. \ref{sec:hamfam}; one of the features that guarantees that the theories analyzed in that section (and also the ones in Sec. \ref{sec:nogravs}) contain only two degrees of freedom is that $\mu=1$ for them [equivalently, the momenta $\pi^{ij}$ enter through the combination $\Pi$ in Eq. \eqref{eq:ham&mom}].

In summary, including potential terms such as $\sqrt{h}\,R_{ij}R^{ij}$ in Eq. \eqref{eq:ric2pot} generally makes the Hamiltonian constraint second class and therefore leaves spatial diffeomorphisms as the only gauge symmetries of the theory. As a consequence, it is not possible to guarantee that the value of $\mu$ is protected against radiative corrections, and therefore that these theories do not acquire additional  degrees of freedom in this way \cite{Sotiriou2017}. Hence, the theory may describe instead two polarizations of the gravitational field with modified dispersion relation, and an additional scalar graviton.\footnote{Note that, even if $\mu=1$ is not protected under radiative corrections, the extra scalar mode may be suppressed in the infrared if $\mu=1$ is an infrared fixed point. Some examples of this behavior, in the framework of Ho\v{r}ava-Lifshitz gravity, are given in \cite{Mukohyama2012,Mukohyama2010}.}

\section{Coupling to matter}\label{sec:matter}

Our analyses above have been restricted to theories which only include the gravitational field. In this last section we include matter fields (for simplicity, a scalar field). This is motivated due to two reasons. The first one is that any realistic theory of gravity must include matter. The second one is that in \cite{Lin2017} the authors study Friedmann-Lema\^itre-Robertson-Walker solution in the square root gravity studied in Sec. \ref{sec:sqrt} and find that the corresponding equations are genuinely different than the ones obtained in general relativity. This may suggest that, even if the square root gravity may be equivalent to general relativity in vacuum, differences might be unavoidable when couplings to matter are included. However, we think that one must be careful before jumping into this conclusion, as many subtleties arise in the procedure of coupling matter in these alternative theories.

\subsection{Algebra of constraints for minimal coupling}

Let us come back to our starting point, namely the Hamiltonian theories introduced in Sec. \ref{sec:hamfam}, and include a scalar field in the system as the simplest possible representation of matter fields. The Lagrangian density for the scalar field is given by
\begin{equation}
\mathcal{L}^\phi=\sqrt{-g}\left[g^{\mu\nu}\nabla_\mu\phi\nabla_\nu\phi+V(\phi)\right],
\end{equation}
where $V(\phi)$ is the potential. In order to obtain the form of the Hamiltonian let us recall that, in the ADM decomposition, the components of the metric are given by
\begin{equation}
g^{00}=-\frac{1}{N^2}\qquad g^{0i}=\frac{N^i}{N^2},\qquad g^{ij}=h^{ij}-\frac{N^iN^j}{N^2}.
\end{equation}
Taking also into account that $\sqrt{-g}=-N\sqrt{h}$, it follows that
\begin{equation}\label{min_coup:ADM}
\mathcal{L}^\phi=-\frac{1}{2}\sqrt{h}\left[-\frac{1}{N}\partial_0\phi\partial_0\phi+2 \frac{N^i}{N}\partial_0\phi\nabla_i\phi+N\,h^{ij}\nabla_i\phi\nabla_j\phi-\frac{N^iN^j}{N}\nabla_i\phi\nabla_j\phi+V(\phi)\right],
\end{equation}
so that the scalar field Hamiltonian is given by ($P$ is the conjugate momentum of $\phi$):
\begin{equation}
H^\phi=\int d^3x\,N\left(\frac{1}{2}\frac{P^2}{\sqrt{h}}+\frac{1}{2}\sqrt{h}h^{ij}\nabla_i\phi\nabla_j\phi+\sqrt{h}V(\phi)\right)
+PN^i\nabla_i\phi.
\end{equation}
If we define
\begin{equation}
\mathcal{H}^\phi=\frac{1}{2}\frac{P^2}{\sqrt{h}}+\frac{1}{2}\sqrt{h}h^{ij}\nabla_i\phi\nabla_j\phi+\sqrt{h}V(\phi),\qquad\qquad \mathcal{D}^\phi_i=P\nabla_i\phi,
\end{equation}
the total (gravity plus matter) Hamiltonian becomes 
\begin{equation}\label{key}
H_{\rm T}=\int d^3x\,N(\mathcal{H}+\mathcal{H}^\phi)+ N^i(\mathcal{D}_{i}+\mathcal{D}^\phi_{i})+ \lambda_N\pi_N+ \lambda^i\pi_{i}.
\end{equation}
The primary constraints are again
\begin{equation}\label{eq:mattprim}
\pi_N=0,\qquad\qquad\pi_{i}=0,
\end{equation}
while the subsequent secondary constraints are given by
\begin{equation}\label{eq:mattsec}
\mathcal{H}_{\rm T}=\mathcal{H}+\mathcal{H}^\phi=0,\qquad\qquad (\mathcal{D}_{\rm T})_i=\mathcal{D}_i+\mathcal{D}^\phi_i=0.
\end{equation}
In order to obtain the off-shell algebra of constraints satisfied, it is convenient to evaluate first the following brackets: first of all the Poisson brackets involving the gravitational and matter sectors of the Hamiltonian constraint,
\begin{align}
&\left\{\mathcal{H}[\alpha],\mathcal{H}[\beta]\right\}=\mathcal{D}[F'^2h^{ij}\left(\alpha\nabla_j\beta-\beta\nabla_j\alpha\right)]\nonumber\\
&\left\{\mathcal{H}[\alpha],\mathcal{H}^\phi[\beta]\right\}=-\int d^3x\,\frac{\lambda}{2\sqrt{h}}\left(\frac{\pi P^2}{\sqrt{h}}-\pi V(\phi) + \sqrt{g}\pi^{ij}\nabla_i\phi\nabla_j\phi \right),\nonumber\\
&\left\{\mathcal{H}^\phi[\alpha],\mathcal{H}^\phi[\beta]\right\}=\mathcal{D}^\phi[h^{ij}\left(\alpha\nabla_j\beta-\beta\nabla_j\alpha\right)].
\end{align}
For the diffeomorphism constraint, one has
\begin{align}
 &\left\{\mathcal{D}[\alpha^i],\mathcal{D}[\beta^j] \right\}=\mathcal{D}\left[\mathcal{L}_{\alpha^i} \beta^j\right],\nonumber\\
 & \left\{\mathcal{D}[\alpha^i],\mathcal{D}^\phi[\beta^j] \right\}=0,\nonumber\\
 &\left\{\mathcal{D}^\phi[\alpha^i],\mathcal{D}^\phi[\beta^j] \right\}=\mathcal{D}^\phi\left[\mathcal{L}_{\alpha^i} \beta^j\right].
\end{align}
Lastly, the brackets involving the different sectors of the Hamiltonian and diffeomorphism constraints are
\begin{align}
&\left\{\mathcal{D}[\alpha^i],\mathcal{H}[\alpha]\right\}=\mathcal{H}[\mathcal{L}_{\alpha^i}\alpha],\nonumber\\
&\left\{\mathcal{D}[\alpha^i],\mathcal{H}^\phi[\alpha]\right\}=\int d^3x\left\{\frac{1}{2}\frac{P^2}{\sqrt{h}}\nabla_i\alpha^i-
\frac{1}{2}\sqrt{h}\nabla_k\phi\nabla^k\phi\nabla_i \alpha^i+\sqrt{h}\nabla_i\phi\nabla_j\phi\nabla^if\alpha^j-\sqrt{h}V(\phi)
\nabla_i \alpha^i\right\},\nonumber\\
&\left\{\mathcal{D}^\phi[\alpha^i],\mathcal{H}[\alpha]\right\}=0,\nonumber\\
&\left\{\mathcal{D}^\phi[\alpha^i],\mathcal{H}^\phi[\alpha]\right\}=\int d^3x \left\{-\frac{P\alpha}{\sqrt{h}}\nabla_i
\left(P \alpha^i\right)+\sqrt{h}\nabla_i\left(\alpha \nabla^i\phi\right) \alpha^j\nabla_j\phi 
-\sqrt{h}\frac{\partial V(\phi)}{\partial \phi}\alpha \alpha^i\nabla_i\phi  \right\}.
\end{align}
Combining these equations, it is now easy to write the off-shell algebra of constraints:
\begin{align}\label{eq:grav+matt}
&\left\{\mathcal{H}_{\rm T}[\alpha],\mathcal{H}_{\rm T}[\beta]\right\}=\mathcal{D}\left[\lambda F'^2(R+\lambda\Pi/h)h^{ij}\left(\beta\partial_j\alpha-\alpha\partial_j\beta\right)\right]+\mathcal{D}^\phi\left[h^{ij}\left(\beta\partial_j\alpha-\alpha\partial_j\beta\right)\right],\nonumber\\
&\left\{\mathcal{D}_{\rm T}[\alpha^i],\mathcal{H}_{\rm T}[\alpha]\right\}=\mathcal{H}_{\rm T}\left[\mathcal{L}_{\alpha^i}\alpha\right],\nonumber\\
&\left\{\mathcal{D}_{\rm T}[\alpha^i],\mathcal{D}_{\rm T}[\beta^j]\right\}=\mathcal{D}_{\rm T}[\mathcal{L}_{\alpha^i} \beta^j].
\end{align}
The only change in the off-shell algebra of constraints \eqref{eq:grav+matt} with respect to the vacuum case appears in the Poisson bracket $\left\{\mathcal{H}_{\rm T}[\alpha],\mathcal{H}_{\rm T}[\beta]\right\}$. That this Poisson bracket is no longer proportional to any of the primary constraints \eqref{eq:mattprim} or secondary constraints \eqref{eq:mattsec} implies the existence of a tertiary constraint
\begin{equation}\label{key}
\lambda F'^2\,\mathcal{D}_i + \mathcal{D}^\phi_i=0.
\end{equation}
This is an independent combination of the two sectors of the diffeomorphism constraint $(\mathcal{D}_{\rm T})_i=0$. Therefore we take as independent constraints 
\begin{equation}
(\mathcal{D}_{\rm T})_i=0,\qquad\mathcal{D}^\phi_i=0.
\end{equation}
The constrain $\mathcal{D}^\phi_i=P\partial_i\phi=0$ does however not satisfy the regularity conditions \cite{Henneaux1994} that are needed in order to proceed with the standard counting of degrees of freedom. As explained in \cite{Henneaux1994}, the matrix obtained by taking the derivatives of the constraint with respect to the variables in the phase space must have constant rank on the constraint surface.

Therefore, we have to choose either $P=0$ or $\partial_i\phi=0$. Let us discuss each of these cases separately:
\begin{itemize}
\item{$P=0$: first of all, consistency with the dynamical evolution implies
\begin{equation}
\{P,H_{\rm T}\}=\{P,\mathcal{H^\phi}[N]+\mathcal{D}^\phi[N^i]\}=\sqrt{h}\partial_i(N\partial^i\phi)-\sqrt{h}N\frac{\partial V}{\partial \phi}=0.
\end{equation}
This leads to a quaternary constraint, the smeared version of which reads
\begin{equation}
\mathcal{C}[\mu]=\int\text{d}^3x\sqrt{h}\,\mu\left[\partial_i(N\partial^i\phi)-N\frac{\partial V}{\partial\phi}\right].
\end{equation}
The complete Hamiltonian is therefore given by
\begin{equation}\label{key}
H_{\rm T}=\int d^3x\,N(\mathcal{H}+\mathcal{H}^\phi)+ N^i(\mathcal{D}_{i}+\mathcal{D}^\phi_{i})+ \lambda_N\pi_N+ \lambda^i\pi_{i}+\lambda_PP+\lambda_{\mathcal{C}}\mathcal{C}.
\end{equation}
From all the constraints, $\mathcal{H}+\mathcal{H}^\phi$, $\pi_N$, $P$ and $\mathcal{C}$ are second-class constraints. This implies that the phase space is 5-dimensional, similarly to what happens in Ho\v{r}ava-Lifshitz gravity \cite{Li2009,Blas2009}. Therefore $\phi$ cannot describe standard matter degrees of freedom.
}
\item{$\phi-f(t)=0$: we impose this form of the constraint (instead of $\partial_i\phi=0$) in order to ensure a more direct matching between the constraints and the degrees of freedom removed. Again, consistency with the dynamical evolution implies
\begin{equation}
-\dot{f}+\{\phi,H_{\rm T}\}=-\dot{f}+\frac{N P}{\sqrt{h}},
\end{equation}
where $\dot{f}=\text{d}f/\text{d}t$. Similarly to the previous situation, these two additional constraints, together with $\mathcal{H}+\mathcal{H}^\phi$ and $\pi_N$, are second-class constraints. The counting of degrees of freedom is therefore the same as in the example just above.
}
\end{itemize}
Summarizing, coupling a scalar field in a minimal way can only be consistent if the scalar field is constrained. Therefore, this scalar field cannot be used to describe standard matter.
Introducing an unconstrained scalar field would only be possible if explicitly breaking the symmetries of the gravitational action, hence exciting additional degrees of freedom (let us note that something similar would happen if a scalar field is coupled to general relativity in a way that diffeomorphism invariance is not preserved.). This result would apply to any kind of field that is minimally coupled. Therefore, in order to compare these theories with general relativity coupled to matter (or to explore their possible phenomenological consequences), first of all one must find how to couple consistently matter to these theories in a way that no constraints have to be satisfied by the matter fields.
In the two next sections we discuss briefly two attempts at achieving this.

\subsection{Two scalar fields}

The first possibility we may consider is adding more scalar fields, and then trying to use the degrees of freedom of one of them in order to ensure that the constraints are satisfied, leaving the others  arbitrary. In this way the constrained scalar field could be interpreted as an additional degree of freedom of gravitational nature, and the other fields would represent the genuine matter sector of the theory.

Let us perform the analysis with two scalar fields (the generic case is an straightforward generalization and does not add much to the discussion):
\begin{equation}
\mathcal{L}^{\phi,\chi}=\sqrt{-g}\left(g^{\mu\nu}\nabla_\mu\phi\nabla_\nu\phi+g^{\mu\nu}\nabla_\mu\chi\nabla_\nu\chi\right).
\end{equation}
The Hamiltonian analysis is completely parallel to the previous discussion, being the Hamiltonian
\begin{equation}
\mathcal{H}^{\phi,\chi}=\mathcal{H}^\phi+\mathcal{H}^\chi=\frac{1}{2}\frac{P_\phi^2}{\sqrt{h}}+\frac{1}{2}\sqrt{h}\,h^{ij}\nabla_i\phi\nabla_j\phi+\frac{1}{2}\frac{P_\chi^2}{\sqrt{h}}+\frac{1}{2}\sqrt{h}\,h^{ij}\nabla_i\chi\nabla_j\chi,
\end{equation}
while the momentum constraints are given by 
\begin{equation}
\mathcal{D}^{\phi,\chi}_i=\mathcal{D}^\phi_i+\mathcal{D}^\chi_i=P\partial_i\phi+ P_\chi\partial_i\chi.
\end{equation}
In this situation, the additional constraints read
\begin{equation}
\mathcal{D}^\phi_i+\mathcal{D}^\chi_i=0.
\end{equation}
If it is possible to satisfy this equation constraining only one field, then the first field could be interpreted as a gravitational degree of freedom, whereas the second one can describe actual (unconstrained) matter. But then, one would also have to impose the consistency of these constraints. As it happens for the single scalar field, this cannot be done in general because the constraints 
\begin{equation}\label{constr}
P_\phi\partial_i\phi+P_\chi\partial_i\chi =0
\end{equation}
do not satisfy the regularity conditions. Instead of using Eq. \eqref{constr} one may consider, for instance, imposing
\begin{equation}\label{constr_2}
\phi=\chi,\qquad\qquad  P_\phi=-P_\chi.
\end{equation}
However, the consistency condition 
\begin{equation}
\{\phi-\chi,H_{\rm T}\}=0
\end{equation}
leads then to the additional constraint
\begin{equation}
P_\phi-P_\chi=0.
\end{equation}
Together with the second constraint in Eq. \eqref{constr_2}, these would imply that $P_\phi=0$ and $P_\chi=0$ identically, so that none of the scalar fields can be unconstrained.

\subsection{Non-minimal coupling}

The previous observations regarding one or several scalar fields are valid if minimal coupling is considered. Let us therefore entertain the possibility that the couplings have to be non-minimal. This is certainly a natural possibility if keeping in mind that these gravitational theories might be obtainable from general relativity by a field redefinition that would transform the minimal couplings into non-minimal ones. Furthermore, the root of the issues illustrated above is the different symmetry groups of the gravitational and matter sector; given that the gravitational theory is not invariant under diffeomorphisms, one should couple matter in a way that is compatible with the different symmetry group of the gravitational sector.

The problem with this approach is that there is apparently no guiding principle that can point to the correct form of the couplings. Therefore, we have considered a number of educated guesses, but none of them have worked. For instance, we have considered a scalar field Lagrangian of the form
\begin{equation}
\mathcal{L}^\phi=\frac{1}{2}\sqrt{-g}\left[\eta_1\,g^{00} \partial_0\phi\partial_0\phi+2\eta_2\,g^{0i}\partial_0\phi\nabla_i\phi+\eta_3\,g^{ij}\nabla_i\phi\nabla_j\phi+V(\phi)\right],
\end{equation}
with $ \eta_1$, $\eta_2 $ and $ \eta_3$ arbitrary functions of the metric and the combination $ R+\lambda\Pi/g$. We have also considered adding to the matter Hamiltonian terms of the form
\begin{equation}
\mathcal{H}^\phi+\phi f\left(R+\lambda\Pi/g\right)+P \tilde{f}\left(R+\lambda\Pi/g\right).
\end{equation}
None of these modifications allows closing the algebra. Overall, the message of this section is just the warning that minimal coupling is not a consistent way to add matter to these theories, and that only very particular non-minimal coupling might do the job, with the caveat that finding the particular form of these couplings seems difficult.
Let us stress that this conclusion is completely general, given that our analysis only relies on the modification of the Hamiltonian constraint which is a characteristic shared by all the theories considered in \cite{Lin2017}.

\section{Conclusions}

In this paper, we have presented a comprehensive analysis of the class of theories found in \cite{Lin2017}, which are characterized by the fact that these contain the same number of degrees of freedom as general relativity. We have used a battery of techniques in order to gain an intuition of the physical meaning of the existence of these theories.

Our main conclusion is that these results strengthen the view that general relativity is a unique theory that is not easily deformed (which is compatible with previous partial analyses \cite{Khoury2011,Khoury2013}). This claim follows from a combination of results:
\begin{itemize}
\item{There are certainly examples that are manifestly different from general relativity (as explained in Sec. \ref{sec:diff_theor}). However, these examples entail profound deviations from general relativity; either these are too trivial to describe gravity and experimentally excluded, or it is not clear if these really have only two degrees of freedom, as it does not seem possible to ensure that the removal of additional degrees of freedom is protected under radiative corrections.}
\item{The most interesting hypothetical theories would exhibit the same linearized properties of general relativity (perhaps at low energy), but would include non-linear deviations from the latter. The formalism developed in \cite{Lin2017} is general enough to encompass these theories, and we have studied in Secs. \ref{sec:hamfam} and \ref{sec:sqrt} quite general families of theories which would seemingly fall into this category. However, every attempt to show differences with respect to general relativity has only demonstrated the existence of unexpected cancellations that eventually dissipate the possible sources of these differences. This has been shown to occur on-shell, both at the linear and non-linear level.}
\end{itemize}
These main conclusions do not imply that the class of theories found in \cite{Lin2017} is not interesting. On the contrary, we think that the results by these authors provide a convenient framework in which a series of new questions about the possibility of deforming general relativity can be formulated. Let us stress that the theories analyzed here escape known results regarding the on-shell constructibility of the scattering amplitudes of general relativity, in particular due to the introduction of higher-derivative interaction vertices. In our opinion, these results represent the strongest argument that general relativity can be obtained recursively from the self-interaction of gravitons in an unambiguous way. Whether a similar statement remains valid in this extended framework remains to be seen, although our results can be considered an important first step in this direction.

A possibility for constructing a counterexample is looking for a theory in which the extrinsic curvature $K_{ij}$ does not only appears in the combination $\mathcal{K}$. A theory of this kind, that solves the conditions given in \cite{Lin2017} in order to have two local degrees of freedom, would most probably avoid our results in Secs. \ref{sec:hamfam} and \ref{sec:sqrt}. This would open the possibility of finding a non-trivial alternative to general relativity. However, it is not clear whether solutions of this form exist.

An additional point that we think is worth keeping in mind, in case one wants to study the observational implications of these theories, is that coupling matter to these theories is a non-trivial task. The underlying reason is that the coupling scheme of matter fields must respect the symmetry group of the gravitational action. Minimal coupling does not work, and standard non-minimal couplings do not seem to be enough in order to guarantee consistency. Hence this feature hinders the application of these theories to phenomenological studies, and its resolution should be further sought in the future.
\section*{Acknowledgments}
The authors would like to thank Matt Visser and Jose Beltr\'an for useful discussions and Shinji Mukohyama and Chunshan Lin for their critical reading of the manuscript. This publication was made possible through the support of the grant from the John Templeton Foundation No.51876. The opinions expressed in this publication are those of the authors and do not necessarily reflect the views of the John Templeton Foundation.
\appendix

\section{Algebra of constraints for an arbitrary $F(x)$ \label{sec:app1}}

Here we make explicit some of the steps in the computation of the algebra \eqref{gen_algebra}.

\subsection{Poisson bracket of the Hamiltonian constraint with itself}

Let us evaluate the bracket
\begin{equation}
\left\{ \mathcal{H}\left[\alpha\right],\mathcal{H}\left[\beta\right]\right\},
\end{equation}
where
\begin{equation}
\mathcal{H}\left[\alpha\right]=\int d^{3}x\,\alpha\mathcal{H}=\int d^{3}x\sqrt{h}\,\alpha F(R+\lambda\Pi/h)
\end{equation}
and $\Pi=\pi^{ij}\pi^{kl}\left(g_{ik}g_{jl}-\frac{1}{2}g_{ij}g_{kl}\right)$; this is the setting introduced in Sec. \ref{sec:hamfam}. We have
\begin{align}
\delta\mathcal{H}[\alpha]&=\sqrt{h}\left\{ F'\left(\nabla^{i}\nabla^{j}\delta g_{ij}-g^{ij}\nabla_{k}\nabla^{k}\delta h_{ij}\right)+...\right\} \alpha\nonumber\\
&+2\sqrt{h}\,\frac{\lambda}{h}\left\{ F'\left(\pi_{ij}-\frac{1}{2}\pi h_{ij}\right)\delta\pi^{ij}\right\} \alpha,
\end{align}
where the dots in the first line correspond to terms that do not contain the derivative of the variation and that will therefore cancel due to the antisymmetrization on $\alpha\leftrightarrow\beta$. Therefore,
\begin{align}
\left\{ \mathcal{H}\left[\alpha\right],\mathcal{H}\left[\beta\right]\right\} &= \int\text{d}^3x\left[2\lambda F'\beta\pi_{ij}\nabla^{i}\nabla^{j}\left(F'\alpha\right)-\alpha\leftrightarrow\beta\right]\nonumber\\
 & =\int\text{d}^3x\left[\left\{-2\lambda\nabla^{i}\left(F'\beta\right)\pi_{ij}\nabla^{j}\left(F'\alpha\right)-2\lambda F'\beta\nabla^{i}\pi_{ij}\nabla^{j}\left(F'\alpha\right)\right\}-\alpha\leftrightarrow\beta\right]\nonumber\\
 & =\int\text{d}^3x\left[\left\{-2\lambda F'\beta\nabla^{i}\pi_{ij}\nabla^{j}F'\alpha-2\lambda F'^{2}\beta\nabla^{i}\pi_{ij}\nabla^{j}\alpha\right\}-\alpha\leftrightarrow\beta\right]\nonumber\\
 & =\int\text{d}^3x\left[-2\lambda F'^{2}\nabla^{i}\pi_{ij}\left(\beta\nabla^{j}\alpha-\alpha\nabla^{j}\beta\right)\right]\nonumber\\
 &=\mathcal{D}\left[\lambda F'^{2}\left(\beta\nabla^{i}\alpha-\alpha\nabla^{i}\beta\right)\right].
\end{align}
%

\subsection{Poisson bracket of the Hamiltonian and diffeomorphism constraints}

Now we want to evaluate the Poisson bracket
\begin{equation}
\left\{ \mathcal{H}\left[\alpha\right],\mathcal{D}\left[\alpha^{i}\right]\right\},
\end{equation}
where the Hamiltonian constraint is the same as before, and the (smeared) diffeomorphism constraints are given by
\begin{equation}
\mathcal{D}\left[\alpha^i\right]=2\int\text{d}^3x\,\pi^{ij}g_{jk}\nabla_{i}\alpha^{k}.
\end{equation}
The variation of the Hamiltonian constraint is given by
\begin{align}
\delta\mathcal{H}[\alpha]&=\frac{1}{2}\int\text{d}^3x\,\mathcal{H}\alpha h^{ij}\delta h_{ij}+\int\text{d}^3x\sqrt{h} \left\{ \frac{\partial F}{\partial R}\left[-R^{ij}\delta h_{ij}+\left(\nabla^{i}\nabla^{j}\delta h_{ij}-h^{ij}\nabla_{k}\nabla^{k}\delta h_{ij}\right)\right]+\right.\nonumber\\
 & \left.+\frac{\lambda}{h}\frac{\partial F}{\partial\left(\frac{\lambda}{h}\Pi\right)}\left[-\Pi h^{ij}\delta h_{ij}+2\left(\pi^{ik}\pi^{j}\,_{k}-\frac{1}{2}\pi^{ij}\pi\right)\delta h_{ij}+2\left(\pi_{ij}-\frac{1}{2}\pi h_{ij}\right)\delta\pi^{ij}\right]\right\} \alpha.
\end{align}
On the other hand, the variation of the diffeomorphism constraint is
\begin{equation}
\delta\mathcal{D}\left[\alpha^{i}\right]=2\int d^{3}x\left[\nabla_{i}\alpha_{j}\delta\pi^{ij}+\pi^{ik}\nabla_{k}\alpha^{j}\delta h_{ij}-\frac{1}{2}\nabla_{k}\left(\pi^{ij}\alpha^{k}\right)\delta h_{ij}\right].
\end{equation}
With these equations, we can evaluate the Poisson bracket
\begin{equation}
\left\{ \mathcal{H}\left[\alpha\right],\mathcal{D}\left[\alpha^{k}\right]\right\} =\int d^{3}x\underbrace{\left(\frac{\delta\mathcal{H}\left[\alpha\right]}{\delta h_{ij}}\frac{\delta \mathcal{D}\left[\alpha^{k}\right]}{\delta\pi^{ij}}\right.}_{\text{I}}-\underbrace{\left.\frac{\delta\mathcal{H}\left[\alpha\right]}{\delta\pi^{ij}}\frac{\delta \mathcal{D}\left[\alpha^{k}\right]}{\delta h_{ij}}\right)}_{\text{II}}.
\end{equation}
Let us now evaluate the two terms marked in the equation above independently:
\begin{itemize}
\item[I.]{This term is given by
\begin{align}\label{eq:apphh1}
& \mathcal{H}\alpha\nabla_{i}\alpha^{i}+\sqrt{h}\left\{ F'\left[-2R^{ij}\nabla_{i}\alpha_{j}-\frac{2\lambda\Pi}{h}\nabla_{i}\alpha^{i}+\frac{2\lambda}{h}\left(\pi^{ik}\pi^{j}\,_{k}-\frac{1}{2}\pi^{ij}\pi\right)2\nabla_{i}\alpha_{j}\right]\alpha+\right.\nonumber\\
 & +\left.2\alpha_{j}\left(-\nabla_{i}\nabla^{i}\nabla^{j}+\nabla^{j}\nabla_{k}\nabla^{k}\right)F'\alpha\right\};
\end{align}
the last term can be rewritten as
\begin{align}
\alpha_j\left(-\nabla_{i}\nabla^{i}\nabla^{j}+\nabla^{j}\nabla_{k}\nabla^{k}\right)F'\alpha&= -\alpha_jR^{j}_{\ i}\nabla^{i}F'\alpha\nonumber\\
& =F'\alpha R^{j}_{\ i}\nabla^{i}\alpha_{j}+\frac{1}{2}F'\alpha^{i}\alpha\,\partial_{i}R,
\end{align}
where in the last identity, a total divergence has been discarded. Hence we can write Eq. \eqref{eq:apphh1} as
\begin{equation}
\mathcal{H}\alpha\nabla_{i}\alpha^{i}+\sqrt{h}\left\{ F'\left[\alpha^{i}\partial_{i}R-\frac{2\lambda\Pi}{h}\nabla_{i}\alpha^{i}+\frac{2\lambda}{h}\left(\pi^{ik}\pi^{j}_{\ k}-\frac{1}{2}\pi^{ij}\pi\right)2\nabla_{i}\alpha_{j}\right]\alpha\right\}.\label{eq:apphh2}
\end{equation}
}
\end{itemize}
\begin{itemize}
\item[II.]{The second piece is simply given by
\begin{equation}
\sqrt{h}\left\{ F'\frac{4\lambda}{h}\left(\pi_{ij}-\frac{1}{2}\pi h_{ij}\right)\left(\pi^{ik}\nabla_{k}\alpha^{j}-\frac{1}{2}\nabla_{k}\left(\pi^{ij}\alpha^{k}\right)\right)\right\} \alpha.\label{eq:apphh3}
\end{equation}
}
\end{itemize}
Combining Eqs. \eqref{eq:apphh2} and \eqref{eq:apphh3} we can write
\begin{align}
\left\{ \mathcal{H}\left[\alpha\right],\mathcal{D}\left[\alpha^{i}\right]\right\} = & \int d^{3}x\,\mathcal{H}\alpha\nabla_{i}\alpha^{i}\nonumber\\
 & +\int d^{3}x\sqrt{h}\,F'\left[\alpha^i\partial_{i}R-\frac{2\lambda\Pi}{h}\nabla_{i}\alpha^{i}+\frac{2\lambda}{h}\left(\pi^{ik}\pi^{j}_{\ k}-\frac{1}{2}\pi^{ij}\pi\right)2\nabla_{i}\alpha_{j}\right.\nonumber\\
 &\left.-\frac{4\lambda}{h}\left(\pi_{ij}-\frac{1}{2}\pi h_{ij}\right)\left(\pi^{ik}\nabla_{k}\alpha^{i}-\frac{1}{2}\nabla_{k}\left(\pi^{ij}\alpha^{k}\right)\right)\right] \nonumber\\
 &=\int\text{d}^3x\,\mathcal{H}\alpha\nabla_{i}f^{i}+\int\text{d}^3x\sqrt{h}\left\{ F'\left(\alpha^i\partial_{i}R+\frac{\lambda}{h}\left(\nabla_{i}\Pi\right)\alpha^{i}\right)\right\} \alpha\nonumber\\
 & =\int d^{3}x\left(\mathcal{H}\alpha\nabla_{i}\alpha^{i}+\alpha^i\alpha\partial_{i}\mathcal{H}\right)\nonumber\\
 &=\int\text{d}^3x\,\alpha\nabla_{i}\left(\mathcal{H}\alpha^{i}\right)=-\int d^{3}x\,\mathcal{H}\alpha^{i}\partial_{i}\alpha\nonumber\\
 & =-\mathcal{H}\left[\mathcal{L}_{\alpha^{i}}\alpha\right].
\end{align}
The bracket between two diffeomorphism constraints is exactly the same as in general relativity, the calculation of which can be found in the literature \cite{Thiemann2008,Bojowald2010}.

\section{Extracting the Poincar\'e algebra}\label{sec:Poinc_calc}

In this appendix we detail the calculations that permit to extract the Poincar\'e algebra from the non-linear algebra of Poisson brackets discussed in Sec. \ref{sec:hamfam}. We will do this calculation from scratch, as it is not easy to find it in the literature. First of all, let us note that the coefficients $\{\delta\mu^\alpha,\delta\omega^{\mu\nu}\}$ in Eq. \eqref{eq:genfun1} are the infinitesimal group parameters associated with the transformations of the Poincar{\'e} group, which act on the spacetime coordinates as
\begin{equation}
\delta x^\alpha=-\left(\delta\mu^\beta P_\beta+\delta\omega^{i0}L_{i0}+\frac{1}{2}\delta\omega^{ij}L_{ij}\right)x^\alpha.\label{eq:inftrans}
\end{equation}
This equation just sets the normalization of the elements of the algebra $P_\mu$ and $L_{\mu\nu}$ (note that greek indices are spacetime indices and latin indices are space indices), which will be needed in order to obtain the corresponding Poisson brackets.

\subsection{Translations with translations}

The simplest transformations are spacetime translations. If we consider $\alpha=\delta\mu^0$, $\alpha^i=0$ and $\beta=0$, $\beta^i=\delta\mu^i$, the only bracket in Eq. \eqref{eq:hdefalg} that may not be trivial is given by
\begin{equation}
\{\mathcal{D}[\delta\mu^i],\mathcal{H}[\delta\mu^0]\}=\mathcal{H}[\mathcal{L}_{\delta\mu^i}\delta\mu^0]=0,
\end{equation}
where the last identity follows from the fact that $\delta\mu^0$ is independent of the position. This implies that the bracket between time and space translations is indeed trivial:
\begin{equation}
\{P_0,P_i\}=0.
\end{equation}
Something similar happens when choosing two time translations, or space translations $\alpha=\beta=0$ and $\alpha^i=\delta^i_j$, $\beta^i=\delta^i_k$ with $j\neq k$. Hence we can write
\begin{equation}
    \{P_\mu,P_\nu\}=0.
\end{equation}
\subsection{Translations with boosts/rotations}

Now let us consider a more interesting case, namely the bracket involving time translations and boosts. This situations corresponds to the choices $\alpha=\delta\mu^0$, $\alpha^i=0$ and $\beta=x_j\delta\omega^{j0}$, $\beta^i=0$. The only non-trivial bracket in Eq. \eqref{eq:hdefalg} is in this case
\begin{equation}
\{\mathcal{H}[\delta\mu^0],\mathcal{H}[x_j\delta\omega^{j0}]\}=\mathcal{D}[\theta\delta\mu^0\partial^i(x_j\delta\omega^{j0})]=\mathcal{D}[\theta\delta\mu^0\omega^{i0}].
\end{equation}
This equation implies that the bracket between the generator of time translations and boosts is proportional to a spatial translation:
\begin{equation}
\{P_0,L_{i0}\}\propto P_i.
\end{equation}
But the very same equation can also be used to fix the proportionality constant: using the normalization of the generators given in Eq. \eqref{eq:inftrans}, one must have (note that the minus sign below comes from the global minus sign in the definition of the infinitesimal transformations)
\begin{equation}
\delta\mu^0\omega^{i0}\{P_0,L_{i0}\}=-\theta\delta\mu^0\omega^{i0}P_i,
\end{equation}
which implies (note that the infinitesimal parameters of the different transformations are arbitrary)
\begin{equation}
\{P_0,L_{i0}\}=-\theta P_i.\label{eq:PL1}
\end{equation}
Let us now consider the following cases in less detail:
\begin{itemize}
\item{$\alpha=\delta\mu^0$, $\alpha^i=0$, $\beta=0$, $\beta^i=x_j\delta\omega^{ji}$: the only non-trivial bracket is
\begin{equation}
\{\mathcal{D}[x_j\delta\omega^{ji}],\mathcal{H}[\delta\mu^0]\}=\mathcal{H}[\mathcal{L}_{x_j\delta\omega^{ji}}\delta\mu^0]=0.
\end{equation}
This implies that the application of time translations and rotations is commutative, namely
\begin{equation}
\{P_0,L_{ij}\}=0.\label{eq:PL2}
\end{equation}
}
\item{$\alpha=0$, $\alpha^i=\delta\alpha^i$, $\beta=x_j\delta\omega^{j0}$, $\beta^i=0$: we have to evaluate
\begin{equation}
\{\mathcal{D}[\delta\mu^i],\mathcal{H}[x_j\delta\omega^{j0}]\}=\mathcal{H}[\mathcal{L}_{\delta\mu^i}x_j\delta\omega^{j0}]=\mathcal{H}[\delta\mu_i\omega^{i0}].
\end{equation}
Hence
\begin{equation}
\{P_i,L_{j0}\}=-\delta_{ij}P_0.\label{eq:PL3}
\end{equation}
}
\item{$\alpha=0$, $\alpha^i=\delta\mu^i$, $\beta=0$, $\beta^i=x_j\delta\omega^{ji}$: we have to evaluate
\begin{equation}
\{\mathcal{D}[\delta\mu^i],\mathcal{D}[x_k\delta\omega^{kj}]\}=\mathcal{D}[\mathcal{L}_{\delta\mu^i}(x_k\delta\omega^{kj})]=\mathcal{D}[\delta\mu_i\delta\omega^{ij}].
\end{equation}
In this case we have to take into account the factor $1/2$ in Eq. \eqref{eq:inftrans}. From the equation just above and \eqref{eq:inftrans} it follows that
\begin{equation}
\delta\mu^i\delta\omega^{kj}\{P_i,L_{kj}\}=2\delta\mu_i\delta\omega^{ji} P_j.
\end{equation}
This last equation, together with the antisymmetric character of $L_{jk}$, leads uniquely to
\begin{equation}
\{P_i,L_{jk}\}=\delta_{ik}P_j-\delta_{ij}P_k.\label{eq:PL4}
\end{equation}
}
\end{itemize}
Due to the explicit splitting between time and space in the Hamiltonian formulation, the Poincar{\'e} algebra appears naturally in a form that is not explicitly covariant. But it is straightforward to write Eqs. \eqref{eq:PL1}, \eqref{eq:PL2}, \eqref{eq:PL3} and \eqref{eq:PL4} in an explicitly covariant form. For $\theta=1$ one obtains
\begin{equation}
\{P_\rho,L_{\mu\nu}\}=\eta_{\nu\rho}P_\mu-\eta_{\mu\rho}P_\nu,
\end{equation}
while for $\theta=-1$ one has
\begin{equation}
\{P_\rho,L_{\mu\nu}\}=\delta_{\nu\rho}P_\mu-\delta_{\mu\rho}P_\nu.
\end{equation}
\subsection{Boosts/rotations with boosts/rotations}
\begin{itemize}
\item{Boosts with boosts; $\alpha=x_i\delta\omega^{i0}$, $N^i=0$, $\beta=x_j\delta\bar{\omega}^{j0}$, $\beta^i=0$:
\begin{align}
\{\mathcal{H}[x_i\delta\omega^{i0}],\mathcal{H}[x_j\delta\bar{\omega}^{j0}]&=\mathcal{D}[\theta x_k\delta\omega^{k0}\delta\bar{\omega}^{i0}-x_j\delta\bar{\omega}^{j0}\delta\omega^i]\nonumber\\
&=\mathcal{D}[\theta x_j(\delta\omega^{j0}\delta\bar{\omega}^{i0}-\delta\bar{\omega}^{j0}\delta\omega^{i0})].
\end{align}
This equation implies that
\begin{equation}
\delta\omega^{i0}\delta\bar{\omega}^{j0}\{L_{i0},L_{j0}\}=-\frac{\theta}{2}L_{ji}(\delta\omega^{j0}\delta\bar{\omega}^{i0}-\delta\bar{\omega}^{j0}\delta\omega^{i0})=-\theta\delta\bar{\omega}^{j0}\delta\omega^{i0}L_{ij},
\end{equation}
namely
\begin{equation}
\{L_{i0},L_{j0}\}=-\theta L_{ij}.\label{eq:LL1}
\end{equation}
}
\item{Boosts with rotations; $\alpha=0$, $\alpha^i=x_j\delta\omega^{ji}$, $\beta=x_k\delta\omega^{k0}$, $\beta^i=0$:
\begin{equation}
\{\mathcal{D}[x_j\delta\omega^{ji}],\mathcal{D}[x_k\delta\omega^{k0}]\}=\mathcal{H}[x_j\delta\omega^{ji}\delta\omega^{k0}\delta_{ik}].
\end{equation}
It follows that
\begin{equation}
\delta\omega^{ji}\delta\omega^{k0}\left\{\frac{1}{2}L_{ji},L_{k0}\right\}=-\delta\omega^{ji}\delta\omega^{k0}\delta_{ik}L_{j0},
\end{equation}
so that
\begin{equation}
\{L_{ij},L_{k0}\}=\delta_{ik}L_{j0}-\delta_{jk}L_{i0}.\label{eq:LL2}
\end{equation}
}
\item{Rotations with rotations; $\alpha=0$, $\alpha^i=x_j\delta\omega^{ji}$, $\beta=0$, $\beta^i=x_k\delta\bar{\omega}^{ki}$:
\begin{align}
\{\mathcal{D}[x_k\delta\omega^{ki}],\mathcal{D}[x_l\delta\bar{\omega}^{lj}]\}&=\mathcal{D}[x_k\delta\omega^{kj}\delta\bar{\omega}^{li}\delta_{jl}-x_l\delta\bar{\omega}^{lj}\delta\omega^{ki}\delta_{jk}]\nonumber\\
&=\mathcal{D}[x_k(\delta\omega^{kj}\delta\bar{\omega}^{li}-\delta\bar{\omega}^{kj}\delta\omega^{li})\delta_{jl}].
\end{align}
It follows that
\begin{align}
\delta\omega^{ki}\delta\bar{\omega}^{lj}\left\{\frac{1}{2}L_{ki},\frac{1}{2}L_{lj}\right\}&=\frac{1}{2}(\delta\bar{\omega}^{kj}\delta\omega^{li}-\delta\omega^{kj}\delta\bar{\omega}^{li})\delta_{jl}L_{ki}\nonumber\\
&=\frac{1}{2}\delta\omega^{ki}\delta\bar{\omega}^{lj}\left(\eta_{jk}L_{li}-\eta_{il}L_{kj}\right).
\end{align}
Then,
\begin{equation}
\{L_{ki},L_{lj}\}=\delta_{jk}L_{li}-\delta_{il}L_{kj}+\delta_{kl}L_{ij}-\delta_{ji}L_{lk}.\label{eq:LL3}
\end{equation}
}
\end{itemize}
It is now straightforward to check that Eqs. \eqref{eq:LL1}, \eqref{eq:LL2} and \eqref{eq:LL3} are equivalent, for $\theta=1$, to
\begin{equation}
\{L_{\mu\nu},L_{\rho\sigma}\}=\eta_{\sigma\mu}L_{\rho\nu}-\eta_{\nu\rho}L_{\mu\sigma}+\eta_{\mu\rho}L_{\nu\sigma}-\eta_{\sigma\nu}L_{\rho\mu}.
\end{equation}
For $\theta=-1$ we just need to replace $\eta_{\mu\nu}\longrightarrow\delta_{\mu\nu}$ in the equation above.

\section{Brief review of on-shell techniques for the evaluation of scattering amplitudes \label{sec:app2}}

Let us introduce some elements of the subject of on-shell scattering amplitudes. This appendix is not intended to be a self-contained review, but rather aims at motivating some basic concepts and results (without proofs) that allow a better understanding of some equations in the main text (and Sec. \ref{sec:higher} in particular). For in-depth discussions of this interesting subject, we refer the reader to the extensive literature, including the monographs \cite{Elvang2015,Elvang2013,Henn2014} in which most of the relevant references can be found (see also \cite{CarRubio2017} for a shorter introduction starting from the basics).

A basic ingredient that is needed for our discussion are the so-called spinor-helicity variables. These represent a more convenient set of variables (instead of the momenta) in order to describe on-shell massless particles. The starting point for the motivation of these variables is the observation that the algebra of the Lorentz group $\mbox{SO}(1,3)$ is equivalent to the algebra of two $\mbox{SU}(2)$ copies, i.e., at the complex algebra level there is an isomorphism
\begin{equation}
\mathfrak{so}(1,3)\simeq \mathfrak{su}(2)\times \mathfrak{su}(2)^*.
\end{equation}
Furthermore, the Lorentz group is homeomorphic to the group of $2\times 2$ unitary matrix with unit determinant $\mbox{SL}(2,\mathbb{C})$. Therefore, to any Lorentz vector we can associate a $2\times 2$ complex matrix via the map
\begin{equation}
p_\mu\longrightarrow p^{\dot{a}b}=p_\mu(\bar{\sigma}^\mu)^{\dot{a}b},
\end{equation} 
where $\bar{\sigma}^\mu=(\mathbbm{1},-\sigma^i)$ [similarly, let us define $\sigma^\mu=(\mathbbm{1},\sigma^i)$]. Then,
\begin{equation}
 \det p^{\dot{a}b}=p_\mu p^\mu.\label{eq:detpp}
\end{equation} 
Representations of the Lorentz group can be found looking for representations of $\mbox{SL}(2,\mathbb{C})$. These can be labeled using the $\mbox{SU}(2)$ representations. The basic object transforming under the fundamental representation $\mathcal{M}$ is a 2-component spinor $ \left|\psi\right]_a$ which transforms as 
\begin{equation}
\left|\psi\right]_a\longrightarrow\mathcal{M}^b\,_a\left|\psi\right]_b.
\end{equation}
On the other hand, for the complex conjugate representation one has the transformation rule
\begin{equation}
\left|\psi\right\rangle_{\dot{a}}\rightarrow\mathcal{M}^{*\dot{b}}\,_{\dot{a}}\left|\psi\right\rangle_{\dot{b}}.
\end{equation}
Dotted indices allow keeping track of the different representations of the $\mbox{SU}(2)$ group, but the notation we are using with angle and square brackets permits us to omit them when convenient. These two representations are
\begin{equation}
\left|\psi\right]\sim\left(1/2,0\right),\qquad\qquad\left|\psi\right\rangle\sim \left(0,1/2\right).
\end{equation}
One can also introduce the additional spinors
\begin{equation}
\left[\psi\right|^a=\epsilon^{ab}\left|\psi\right]_b,\qquad\qquad\left\langle\psi\right|^{\dot{a}}=\epsilon^{\dot{a}\dot{b}}\left|\psi\right\rangle_{\dot{b}},
\end{equation}
where
\begin{equation}
\epsilon^{ab}=
\begin{pmatrix}
0  & 1 \\
-1 & 0
\end{pmatrix}=\epsilon^{\dot{a}\dot{b}}.
\end{equation}
With these elements, the Dirac equation can be rewritten exploiting the identity\footnote{The convention used here for the gamma matrices is \begin{equation}
\gamma^\mu=
\begin{pmatrix}
0  				 				 &  (\sigma^\mu)_{a\dot{b}}\\
(\bar{\sigma}^\mu)^{\dot{a}b}	 &	0
\end{pmatrix}.
\end{equation}
}
\begin{equation}
\slashed{p}=
\begin{pmatrix}
0  				 				 &  p_{a\dot{b}}\\
p^{\dot{a}b}	 &	0
\end{pmatrix},
\end{equation}
so that $\slashed{p}=0$ is equivalent to the Weyl equations
\begin{equation}
p^{\dot{a}b}\left|p\right]_b=0,\qquad p_{a\dot{b}}\left|p\right\rangle^{\dot{b}}=0,\qquad \left[p\right|^b p_{b\dot{a}}=0,\qquad \left\langle p\right|_{\dot{b}}p^{\dot{b}a}=0.
\end{equation}
If we take into account Eq. \eqref{eq:detpp}, the determinant of these matrices vanishes identically for on-shell massless particles. It follows that one can write
\begin{equation}
p_{a\dot{b}}=-|p]_a\langle p|_{\dot{b}},\qquad\qquad p^{\dot{a}b}=-|p\rangle^{\dot{a}}[p|^b.\label{eq:shdecomp}
\end{equation}
This is one of the main properties that justifies the usefulness of the spinor-helicity variables: we can use the angle and square brackets instead of the momenta in order to write equations that involve the momentum of on-shell massless particles (e.g., scattering particles); note that this is independent of the helicity of these particles. Not only the momentum of an on-shell massless particle can be written in this way, but also other quantities of interest such as polarization vectors of spin-1 particles, namely
\begin{equation}
\epsilon^\mu_+(p,q)=-\frac{\langle q|\gamma^\mu|p]}{\sqrt{2}\langle qp\rangle},\label{eq:polvec+}
\end{equation}
which corresponds to the polarization with positive helicity, and the corresponding quantity with negative helicity,
\begin{equation}
\epsilon^\mu_-(p,q)=-\frac{\langle p|\gamma^\mu|q]}{\sqrt{2}[qp]}.\label{eq:polvec-}
\end{equation}
Eqs. \eqref{eq:polvec+} and \eqref{eq:polvec-} are just a particular representation of the usual polarization vectors of photons and gluons in terms of spinor-helicity variables (see, e.g., \cite{Elvang2013} for a very explicit discussion of this point). In these equations, $q$ is a reference spinor that does not have any physical meaning and can be chosen arbitrarily, representing the freedom of choosing a particular gauge.

This construction goes on for particles with higher spin. In particular, the polarization tensors describing the two helicities of the gravitational field (which will be the same for all the theories in which the Fierz-Pauli field equations are recovered at linear order)  are given by 
\begin{equation}
\epsilon_\pm^{\mu\nu}=\epsilon^\mu_\pm\epsilon^\nu_\pm.
\end{equation}

These are essentially the main ingredients that are needed for our discussion, together with the use of complexifications of the external momenta. This kind of complexification permits to exploit the analytical structures of scattering amplitudes, which at tree level are rational functions that can only display poles when internal propagating particles (in our case, gravitons) become on-shell; this is what is usually defined as the locality condition. Note that additional poles that do not correspond to physical particles (i.e., the propagating degrees of freedom) may appear in individual Feynman diagrams, but these must cancel in the complete amplitude (these are known as spurious poles). In particular, for the theories analyzed in Sec. \ref{sec:sqrt} we know that the only possible complex poles in the amplitudes correspond to physical gravitons on-shell, i.e., satisfying $\hat{p}^2=0$.

A generic complexification has the form
\begin{equation}
p_s\longrightarrow \hat{p}_s=p_s+zq_s,
\end{equation}
where $s$ marks certain subset of the external legs, $z\in\mathbb{C}$, and $\sum_s q_s=0$ in order to guarantee that momentum conservation holds for the complexified momenta. This defines a complexified $n$-point amplitude $A_n(z)$, such that $A_n(0)$ is the original amplitude evaluated on real momenta. If the $\{q_s\}$ are all orthogonal (which in particular implies that $q_s$ is null) and $p_s\cdot q_s=0$ (with no summation on the index $s$), then the poles in $A_n(z)$ are simple poles. It is then possible to show that, if $A_n(z)\rightarrow0$ when $|z|\rightarrow\infty$, it is possible to construct $A_{n}(0)$ as a sum of quadratic products of on-shell $n$-point amplitudes evaluated on complex momenta. That is, it is possible to determine completely $A_{n}(0)$ from the knowledge of $A_{n-1}(z)$, implying that scattering amplitudes can be constructed recursively.

A sufficient condition to proof that $A_n(z)$ vanishes at infinity is showing that the individual contributions from all the possible Feynman diagrams contributing to a given amplitude vanish independently. In order to show this, one needs to extract the leading behavior with $|z|$ of all the relevant elements in these diagrams. In particular, it will be useful for our arguments in Sec. \ref{sec:higher} to recall some basic properties of Feynman diagrams with $n$ external particles and $k$-point vertices, namely the number of vertices $v$, propagators $p$ and internal vertices $i$ (not stitched to external particles):
\begin{itemize}
\item{\emph{3-point vertices:} $v=n-2$, $p=n-3$ and $i\leq n/2-2$.}
\item{\emph{4-point vertices:} $v=n/2-1$, $p=n/2-2$ and $i\leq n/6-1$.}
\end{itemize}
Note that the number of internal vertices $i$ depends on the topology of the particular diagrams considered. We will not need this information for $k\geq 5$, as the corresponding contributions are subleading (which is typically also true for 4-point vertices with respect to 3-point vertices; for instance, considering only vertices quadratic in the momenta, the corresponding powers coming from the vertices are $n-2$ and $2n-4$, respectively). In general, the number of vertices and propagators as functions of $k$ are given by
\begin{equation}
v(k)=\frac{n-2}{k-2},\qquad p(k)=v(k)-1.
\end{equation}
In the case of general relativity, and the gravitational theories discussed in this paper, the optimal choice  \cite{Benincasa2007,Hall2008} for the complexification ($N>n/2$ is the number of particles with positive helicity) is given by
\begin{equation}
\hat{p}_s=p_s+z|t\rangle[s|,\qquad \hat{p}_t=p_t-z\sum_{s=1}^N|t\rangle[s|.
\end{equation}
where $s\in[1,N]$ labels all the external legs with positive helicity, and $t$ marks a single external leg with negative helicity. In this equation, we have exploited the fact that $\{q_s\}_{s=1}^N$ and $q_t$ are null vectors as well by definition, so that these can be decomposed in spinor-helicity variables as in Eq. \eqref{eq:shdecomp}. If we decompose in this way the momenta (such that, e.g., $p_s=-|s\rangle[s|$), we can directly write this complexification in terms of spinor-helicity variables as
\begin{equation}
|\hat{s}\rangle=|s\rangle-z|t\rangle,\qquad |\hat{t}]=|t]+z\sum_{s=1}^N|s].
\end{equation}
%

\bibliographystyle{unsrt}
\bibliography{refs}

\end{document}